\documentclass[aps,prb,twocolumn,showpacs,superscriptaddress,floatfix]{revtex4-1}
\usepackage{dcolumn}
\usepackage{amsmath}
\usepackage{graphicx,epsfig,psfrag}
\usepackage{amssymb}
\usepackage{natbib}
\usepackage{url}
\usepackage[breaklinks=true]{hyperref}
\usepackage{mathtools}
\usepackage{subfigure}
\hypersetup{
        colorlinks=true,
}
\usepackage{bookmark}
\usepackage{epic}

\usepackage{txfonts}
\usepackage{color}

\begin{document}
\title{Magnetic field dependence of the thermopower of Kondo-correlated quantum dots: Comparison with experiment}
\author{T. A. Costi}
\affiliation 
{Peter Gr\"{u}nberg Institut and Institute for Advanced Simulation, 
Research Centre J\"ulich, 52425 J\"ulich, Germany}
\begin{abstract}
  Signatures of the Kondo effect in the electrical conductance of strongly correlated quantum dots are well understood both experimentally and theoretically, while those in the thermopower have been the subject of recent interest, both theoretically and experimentally. Here, we extend theoretical work [T. A. Costi, Phys. Rev. B {\bf 100}, 161106(R) (2019)] on the field-dependent thermopower of such systems to the mixed valence and empty orbital regimes, and carry out calculations in order to address a recent experiment on the field dependent thermoelectric response of Kondo-correlated quantum dots [A. Svilans {\em et al.,} Phys. Rev. Lett. {\bf 121}, 206801 (2018)]. In addition to the sign changes in the thermopower at temperatures
  $T_1(B)$ and $T_2(B)$ (present also for $B=0$) in the Kondo regime, an additional sign change was found [T. A. Costi, Phys. Rev.  B {\bf 100}, 161106(R) (2019)] at a temperature $T_0(B)<T_1(B)<T_2(B)$ for fields exceeding a gate-voltage dependent value $B_0$, where $B_0$ is comparable to, but larger, than the field $B_c$ at which the Kondo resonance splits. We describe the evolution of the Kondo-induced sign changes in the thermopower at temperatures $T_0(B),T_1(B)$ and $T_2(B)$ with magnetic field and gate voltage from the Kondo regime to the mixed valence and empty orbital regimes and show that these temperatures merge to the single temperature $T_0(B)$ upon entry into the mixed valence regime. By carrying out detailed numerical renormalization group calculations for the above quantities, using appropriate experimental parameters,
  we address a recent experiment which measures the field-dependent thermoelectric response of InAs quantum dots exhibiting the Kondo effect  [A. Svilans {\em et al.,} Phys. Rev. Lett. {\bf 121}, 206801 (2018)]. This allows us to understand
  the overall trends in the measured field- and temperature-dependent thermoelectric response as a function of gate voltage.
In addition, we determine which signatures of the Kondo effect (sign changes at $T_0(B),T_1(B)$ and $T_2(B)$) have been observed in this experiment, and find that while the Kondo-induced signature at $T_1(B)$ is indeed measured in the data,
the signature at $T_0(B)$ can only be observed by carrying out further measurements at a lower temperature. In addition, the less interesting (high-temperature) signature at $T_2(B)\gtrsim \Gamma$, where $\Gamma$ is the electron tunneling rate onto the dot, is found to lie above the highest temperature in the experiment, and was therefore not accessed. Our calculations provide a useful framework for interpreting future experiments on direct measurements of the thermopower of  Kondo-correlated quantum dots in the presence of finite magnetic fields, e.g., by extending zero-field measurements
of the thermopower [B. Dutta {\em et al.,} Nano Lett. {\bf 19}, 506 (2019)] to finite magnetic fields.
\end{abstract}
\maketitle
\section{Introduction}
The Kondo effect, originally describing the anomalous low-temperature increase in the resistivity of nonmagnetic metals due to the presence of magnetic impurities \cite{Kondo1964,Hewson1997} , is by now a ubiquitous phenomenon in condensed matter physics \cite{Cox1998}.  It plays a role, for example, in the decoherence of qubits coupled ohmically to an environment \cite{Leggett1987,Weiss2008,Nghiem2016}, in transition metal atoms in nanowires \cite{Lucignano2009}, in magnetic adatoms on surfaces \cite{Li1998,Madhavan1998,Manoharan2000,Nagaoka2002,Ternes2017,Gruber2018}, in semiconductor \cite{Goldhaber1998b,Cronenwett1998,Schmid1998,vanderWiel2000,Kretinin2011} and molecular quantum dots \cite{Park2002,Yu2005,Scott2010}, in heavy fermions \cite{Hewson1997,Loehneysen2007} and in the Mott transition in strongly correlated materials \cite{Kotliar2004}.

Recently, the Kondo effect has attracted attention in the context of the thermoelectric response of gate-tunable semiconductor and molecular quantum dots, both experimentally \cite{Scheibner2005,Svilans2018,Dutta2019}
and theoretically \cite{Sakano2007,Costi2010, Sakano2007,Weymann2013,Andergassen2011a,Rejec2012,Costi2019a,Karki2019}. Understanding the thermoelectric
properties of such systems is important for using nanoscale thermoelectric elements to improve the energy efficiency
of microelectronic devices \cite{Mahan1998,Sothmann2014,Zimbovskaya2016,Thoss2018}. By comparison with electrical
conductance measurements, however, measurements of the thermopower (Seebeck coefficient), are more challenging \cite{Scheibner2005,Reddy2007,Cui2017,Prete2019,Gehring2019}. Recent works have nevertheless made progress in this direction and some of the predicted signatures of the Kondo effect in the thermopower of strongly correlated
quantum dots \cite{Costi2010,Costi2019a} have been observed \cite{Svilans2018,Dutta2019}. While the electrical conductance $G(T)$ measures the zeroth moment of
the spectral function and is therefore enhanced by the build up of the Kondo resonance with decreasing temperature \cite{Glazman1988,Ng1988,Costi1994,Pustilnik2004}, the thermopower $S(T)$ measures the first moment of the spectral function, which has both positive and negative contributions from a region of width $2k_{\rm B}T$ about the Fermi level \cite{Costi2010}. Thus, sign changes in the thermopower give information about the relative importance of electronlike and holelike contributions to the Kondo resonance and how these depend on temperature and magnetic field.
While previous work exists on the magnetoconductance of Kondo-correlated quantum dots \cite{Costi2001,Hofstetter2001,Karrasch2006}, and for the zero-field thermopower \cite{Costi2010}, only recently has the thermopower in a magnetic field been fully clarified \cite{Costi2019a}.

In this paper, motivated by a recent experimental study of the thermoelectric response of Kondo-correlated quantum dots in the presence of a magnetic field \cite{Svilans2018}, we compare numerical renormalization group (NRG) predictions for the Kondo-induced sign changes in the thermopower at finite magnetic field with experiment. While
Ref.~\onlinecite{Costi2019a} addressed the Kondo-induced sign changes in the slope of the thermopower with respect to gate voltage at midvalley (i.e., at the particle-hole symmetric point of the Anderson model) as a function of field and temperature, in this paper we address these sign changes over the full gate-voltage dependence of the thermopower. We also present the results for the thermopower in a magnetic field in the mixed valence and empty orbital regimes, which were not discussed in Ref.~\onlinecite{Costi2019a}.

The paper is organized as follows. In Sec.~\ref{sec:model+transport} we describe the model for a strongly correlated quantum dot, outline the transport calculations for the thermopower within the NRG approach, define the Kondo scales used in the paper, and outline the different parameter regimes of the model relevant for quantum dots
(Kondo, mixed valence, and empty orbital regimes).
Section~\ref{sec:sign-changes} describes the Kondo-induced
sign changes in the thermopower at temperatures $T_{i=0,1,2}(B)$ in the presence of a magnetic field and the evolution of these with magnetic field and gate voltages ranging from the Kondo regime to the mixed valence regime. The signatures of the Kondo-induced sign changes in the gate-voltage dependence of the thermopower (at selected fixed temperatures and magnetic fields) is described in Sec.~\ref{sec:gate-dependence} for $U/\Gamma \gg 1$, while in Sec.~\ref{sec:comparison} we use the experimental value for $U/\Gamma=3.2$ from Ref.~\onlinecite{Svilans2018} in order to make a comparison between the calculated gate-voltage dependence of the linear-response thermocurrent ($\propto$ thermopower) and the measured gate-voltage dependence of the thermocurrent in Ref.~\onlinecite{Svilans2018} for the same fields and temperatures as in the experiment. Conclusions are given in Sec.~\ref{sec:conclusions}, where we also suggest some directions for future studies. 
Details of the magnetic field dependence of the thermopower in the mixed valence and empty orbital regimes are given in Appendix~\ref{sec-mixed-valence}, while further results for quantum dots with several different values of $U/\Gamma$ are given in Appendices~\ref{sec:table} and \ref{sec:B1Gamma}. Appendix~\ref{sec:GS-comparison} compares the linear-response thermocurrent to the thermopower for the parameters of the experiment \cite{Svilans2018}.

\section{Model and transport calculations}
\label{sec:model+transport}
We describe the thermoelectric transport through a strongly correlated quantum dot within a two-lead single level Anderson impurity model consisting of three terms,
\begin{align}
  H &=H_{\rm dot} + H_{\rm leads} + H_{\rm tunneling}.\label{eq:model}
      \end{align}
      Here, the first term, describing the quantum dot is given by
\begin{align}
  H_{\rm dot} &=\sum_{\sigma}\varepsilon_{0}n_{0\sigma}
  -g\mu_{\rm B}BS_{z}+Un_{0\uparrow}n_{0\downarrow},\label{eq:model-dot}
\end{align}
where $\varepsilon_{0}$ is the level energy, $U$ the 
local Coulomb repulsion $U$, $B$ is a local magnetic field, and  $S_{z}=\frac{1}{2}(n_{0\uparrow}-n_{0\downarrow})$ is the $z$ compoent of the local electron spin.
The second term $H_{\rm leads}$, given by
\begin{align}
H_{\rm leads} &=\sum_{k\alpha=L,R\sigma}(\epsilon_{k\alpha}-\mu_{\alpha})c^{\dagger}_{k\alpha\sigma}c_{k\alpha\sigma},
\label{eq:leads}
\end{align}
describes the two noninteracting conduction electron leads ($\alpha=L,R$), with kinetic energies $\epsilon_{k\alpha}$ and chemical potentials $\mu_{\alpha=L,R}={\epsilon_{\rm F}\pm eV_{\rm bias}}/2$ with  $V_{\rm bias}$ being the bias voltage across the quantum dot. Since we shall only be concerned with linear-response, the limit $V_{\rm bias}\to 0$ is to be understood. Finally,  the last term
\begin{align}
  H_{\rm tunneling}&=\sum_{k\alpha\sigma} t_{\alpha}(c^{\dagger}_{k\alpha\sigma}d_{\sigma}+d^{\dagger}_{\sigma}c_{k\alpha\sigma}),
\end{align}
describes the tunneling of electrons from the leads to the dot with amplitudes $t_{\alpha=L,R}$. In the above, $n_{0\sigma}=d_{\sigma}^{\dagger}d_{\sigma}$ is the number operator for electrons on the dot, $d_{\sigma}^{\dagger}$ ($d_{\sigma}$) and
$c_{k\alpha\sigma}^{\dagger}$ ($c_{k\alpha\sigma}$ ) are electron creation (annihilation) operators,
and we assume 
a constant density of states, $\rho_{\alpha}(\omega)=\sum_{k}\delta(\omega-\varepsilon_{k\alpha})=1/(2D) \equiv N_{\rm F}$ for both leads, with $D=1$ the half-bandwidth and we have set the Fermi level of the leads as our zero of energy, i.e., $\epsilon_{\rm F}=0$.
The strength of correlations is characterized by $U/\Gamma$, where $\Gamma=2\pi N_{\rm F}(t_L^2 +t_R^2)$ is the tunneling rate, taken throughout as $\Gamma=0.002D$. Investigation of the Kondo effect, requires, in general the use
of non-perturbative methods \cite{Wilson1975,KWW1980a,Gonzalez-Buxton1998,Bulla2008,Tsvelick1983b,Andrei2013,Gull2011b,White1992}.
Here, we solve $H$ using the NRG technique \cite{Wilson1975,KWW1980a,Gonzalez-Buxton1998,Bulla2008,Zitko2009b},
which, as we shall describe below, is particularly well suited to the calculation of transport properties.
Since we are primarily interested in interpreting the experiment in Ref.~\onlinecite{Svilans2018}, most
calculations will be for the experimentally determined value of $U/\Gamma=3.2$ \cite{Svilans2018}. We further note that by working in a basis of conduction electron states with well-defined even and odd parities, that only the even-parity combination couples to the impurity, with strength $t=\sqrt{t_{L}^2+t_{R}^2}$, thereby making the NRG calculations reported below effectively single-channel ones.

We define a dimensionless gate voltage ${\rm v}_g\equiv -(\varepsilon_{0}+U/2)/\Gamma$,
such that the particle-hole symmetric (or midvalley) point at $\varepsilon_{0}=-U/2$, where $n_{0}=\sum_{\sigma}n_{0\sigma}=1$, occurs at ${\rm v}_g=0$. The present definition of ${\rm v}_g$,
differing by a minus sign from that used in Ref.~\onlinecite{Costi2019a}, is convenient since
the experimental gate-voltage, $V_g$,  given by $-|e|V_{g} = \varepsilon_0 \sim  - {\rm v}_g\Gamma$, then
has the same sign as ${\rm v}_g$, which facilitates the comparisons with experiment to be shown later.

The linear-response thermopower $S(T)=-I_{1}/|e|TI_{0}$ \cite{Kim2002,Dong2002,Costi2010}, with $e$ the electron charge, is calculated by evaluating the transport integrals
\begin{align}
  I_{m=0,1}&=\gamma\int_{-\infty}^{+\infty}d\omega (-\partial f/\partial \omega) \omega^m A(\omega,T),
             \label{eq:transport-integrals}
\end{align}
where $\gamma=\pi\Gamma/2h$, $h$ is Planck's constant, and $A(\omega,T)=\sum_{\sigma}A_{\sigma}(\omega,T)$, with $A_{\sigma}(\omega,T)$ the spin-resolved local level spectral function of the dot. The latter can be written within a Lehmann representation as
\begin{align}
A_{\sigma}(\omega,T) = &\frac{1}{Z(T)}\sum_{m,n}|\langle m|d_{\sigma}|n\rangle|^2(e^{-E_m/k_{\rm B}T}+e^{-E_n/k_{\rm B}T})\nonumber\\
&\times\delta(E - (E_{n}-E_{m})),\label{eq:spectral-function}
\end{align}
where $E_m$ are NRG eigenvalues and $|m\rangle$ are NRG eigenstates of $H$ and
$Z(T)=\sum_{m}e^{-E_m/k_{\rm B}T}$ is the partition function at temperature $T$.

We follow the approach of Ref.~\onlinecite{Yoshida2009} and evaluate $I_0(T)$ and $I_1(T)$ by inserting the discrete form of the spectral function (\ref{eq:spectral-function}) into Eq.~(\ref{eq:transport-integrals}) to obtain
\begin{equation}
I_{i=0,1}(T) = \frac{\gamma}{k_{\rm B}TZ(T)}\sum_{m,n,\sigma}(E_n-E_m)^{i}\frac{|\langle m|d_{\sigma}|n\rangle|^2}
{(e^{E_m/k_{\rm B}T}+e^{E_n/k_{\rm B}T})}.\label{eq:moments}
\end{equation}
This way of calculating $I_0(T)$ and $I_1(T)$ avoids any additional errors that can arise by first broadening the
spectral function in (\ref{eq:spectral-function}) and then using the resulting smooth spectral functions to carry out explicitly the integrations in (\ref{eq:transport-integrals}). Moreover,
since the expressions for $I_{i=0,1}$ in Eq.~(\ref{eq:moments}) take the same form as those for the calculation of thermodynamic observables within the NRG \cite{KWW1980a,KWW1980b}, and, since the latter are known to be essentially exact by comparisons with thermodynamic Bethe-ansatz calculations\cite{Merker2012b,Merker2013}, the calculations for $S(T)$  (and also the conductance $G(T)$ which follows from $I_0(T)$) are also essentially exact at all
temperatures and for all parameter values (magnetic field, Coulomb repulsion, local level position, etc). We use a logarithmic discretization parameter of $\Lambda=4$ throughout and suppress any induced oscillations in physical quantities at low temperature by using $z$ averaging with $N_z=4$ bath realizations \cite{Oliveira1994,Campo2005}.

By particle-hole symmetry, 
\begin{align}
S_{-{\rm v}_g}(T) & =-S_{+{\rm v}_g}(T),\label{eq:S-ph}
\end{align}
so in describing the gate-voltage dependence of the thermopower, it suffices to consider either
${\rm v}_g<0$ or ${\rm v}_g>0$. We shall mostly consider the former.

Apart from the scale $\Gamma$, we shall also make some use of the Kondo scale, $T_{\rm K}$, 
defined in terms of the $T=B=0$ local spin susceptibility $\chi_0$ via
\begin{align}
  \chi_0&=\frac{(g\mu_{\rm B})^2}{4k_{\rm B} T_{\rm K}},\label{eq:tk-spin}
\end{align}
where $\chi_0$ is evaluated within NRG via
\begin{align}
  \chi_0=\lim_{T\to 0}(g\mu_{\rm B})^2\int_{0}^{1/{k_{\rm B}T}}d\tau \langle S_{z}(\tau)S_{z}(0)\rangle.\label{eq:chi0-matsubara}
  \end{align}

The Kondo scale, $T_{\rm K}$, so defined is comparable to another frequently used Kondo scale, $T_{\rm K1}$, from perturbative scaling \cite{Haldane1978,Hewson1993}, which is given by 
\begin{align}
  \frac{k_{\rm B}T_{\rm K1}({\rm v}_g)}{\Gamma} = & \sqrt{\frac{U}{4\Gamma}}e^{-\pi |\varepsilon_{0}||\varepsilon_{0}+U|/\Gamma U}
= \sqrt{\frac{u}{4}}e^{-\pi(u^2/4-{\rm v}_{g}^2)/u},\label{eq:tk-vs-vg}
\end{align}
where $u=U/\Gamma$. A comparison between these two definitions of the Kondo scale for 
${\rm v}_g=0$ and different values of $U/\Gamma$ is given in Table~\ref{Table2} of Appendix~\ref{sec:table}. 

For strong correlations, i.e., for $U/\Gamma\gg 1$, three regimes can be defined for the Anderson model 
given by Eq.~\ref{eq:model}\cite{KWW1980a,KWW1980b,Hewson1997}: the Kondo regime, when local spin fluctuations predominate, the mixed valence regime, when charge fluctuations are important, and the empty orbital regime, when neither spin nor charge fluctuations are significant and the physics is that of a noninteracting resonant level with only thermal fluctuations playing a significant role. Clearly, the different regimes are adiabatically connected to each other so different definitions, in terms of model parameters, are possible. An approximate definition, in terms of the 
 range of local level positions, is as follows: the Kondo regime, may be approximately defined by
local level positions $-U/2 \leq \varepsilon_0\leq -\Gamma$ (corresponding to dot occupancies satisfying {\em approximately} $0.75\leq n_{0}\leq 1$), and, using particle-hole symmetry, $-U/2 \leq -(\varepsilon_0+U) \leq -\Gamma$ (corresponding to dot occupancies satisfying {\em approximately} $1 \leq n_{0}\leq 1.25$). In terms of the dimensionless gate voltage 
${\rm v}_g\equiv -(\varepsilon_0+U/2)/\Gamma$, and the dimensionless 
charging energy $u=U/\Gamma$, the above range of local levels corresponds to $|{\rm v}_g|\leq (u-1)/2$.
In this regime, the occupancy of the dot lies approximately in the range $0.75 \leq n_0 \leq 1.25$ \cite{Costi2010}.
The mixed valence regime borders on the Kondo regime and may be  defined approximately by local level  
positions $-\Gamma/2 \leq \varepsilon_0\leq +\Gamma/2$, corresponding to $-(u-1)/2 \leq {\rm v}_g \leq -(u + 1)/2$.  In this regime, the charge on the dot fluctuates between $n_0=0$ and 
$1$, and its average value, depending on the precise value of $\varepsilon_0$, can lie 
anywhere in the {\em approximate} range $0.25 \leq n_0\leq 0.75$. 
Another mixed valence regime occurs for $-\Gamma/2 \leq 
\varepsilon_0+U\leq +\Gamma/2$, corresponding to 
dimensionless gate-voltages in the range $+(u-1)/2 \leq {\rm v}_g \leq 
+(u + 1)/2$ and a dot occupancy of around $n_0=1.5$ (lying {\em approximately} in the range 
$1.25\leq n_0\leq 1.75$ depending on the precise value of 
$\varepsilon_0$). Finally, the empty orbital regime with $n_0\approx 0$ is given by 
$\varepsilon_0>\Gamma/2$, i.e., ${\rm v}_g<-(1+u)/2 $ with a similar (full 
orbital) regime at $\varepsilon_0+U<-\Gamma/2$, 
i.e., ${\rm v}_g>+(1+u)/2$, where $n_0\approx 2$. 
While the above can be used as working definitions for the various regimes, the
boundaries between the regimes are not sharp. In particular, for local level positions $\varepsilon_0$ approaching the mixed valence boundary from the Kondo side, significant charge fluctuations will modify some of
the generic features encountered in the Kondo regime.  We shall refer to this narrow range of
level positions $\varepsilon_0$ (of width $\Delta E$) as the ``weak Kondo regime'', i.e.,
$-\Gamma/2 - \Delta E \leq \varepsilon_0 \leq -\Gamma/2$. We find, for $U/\Gamma=3.2$, for
example, that $\Delta E\approx 0.1\Gamma$, so this regime occurs for
$-0.6\Gamma \leq \varepsilon_0 \leq -0.5\Gamma$ (i.e., $0.9 \lesssim {\rm v}_g\lesssim 1.0$ and $-1.0 \lesssim {\rm v}_g\lesssim -0.9$). Taking as an example the case
of the experiment in Ref.~\onlinecite{Svilans2018} with $U/\Gamma=u=3.2$ we find that the Kondo regime occurs for $|{\rm v}_g|\leq (u-1)/2=1.1$, the mixed valence regime occurs for $1.1\leq {\rm v}_g\leq 2.1$ or $-2.1\leq {\rm v}_g\leq -1.1$ and the empty (full) orbital regime occurs for $|{\rm v}_g|>2.1$. In contrast, if we use as criterion for the different regimes that the dot occupancy lies exactly within the above given ranges, then we find that the Kondo regime occurs for $|{\rm v}_g|\leq 1.0$, the mixed valence regime occurs for  $1.0\leq {\rm v}_g\leq 2.4$ (and $-2.4\leq {\rm v}_g\leq -1.0$) and the the empty (full) orbital regime occurs for $|{\rm v}_g|>2.4$. While the former definition is simpler, we shall use the latter in the calculations relating to the experiment : the main effect is that the Kondo regime is delineated from the mixed valence regime by $|{\rm v}_g|\leq 1.0$ instead of $|{\rm v}_g|\leq 1.1$.

Unless otherwise stated, we shall henceforth set the $g$ factor $g$, the Bohr magneton $\mu_{\rm B}$, the Boltzmann constant $k_{\rm B}$, the electric charge $e$, and, Planck's constant $h$ to unity throughout 
($g=\mu_{\rm B}=k_{\rm B}=|e| = h =1$). Hence, expressions such as $T/\Gamma$ and $B/\Gamma$ should be read as $k_{\rm B}T/\Gamma$ and $g\mu_{\rm B}/\Gamma$, respectively. 
\begin{figure*}[t]
\centering 
\includegraphics[width=0.95\columnwidth]{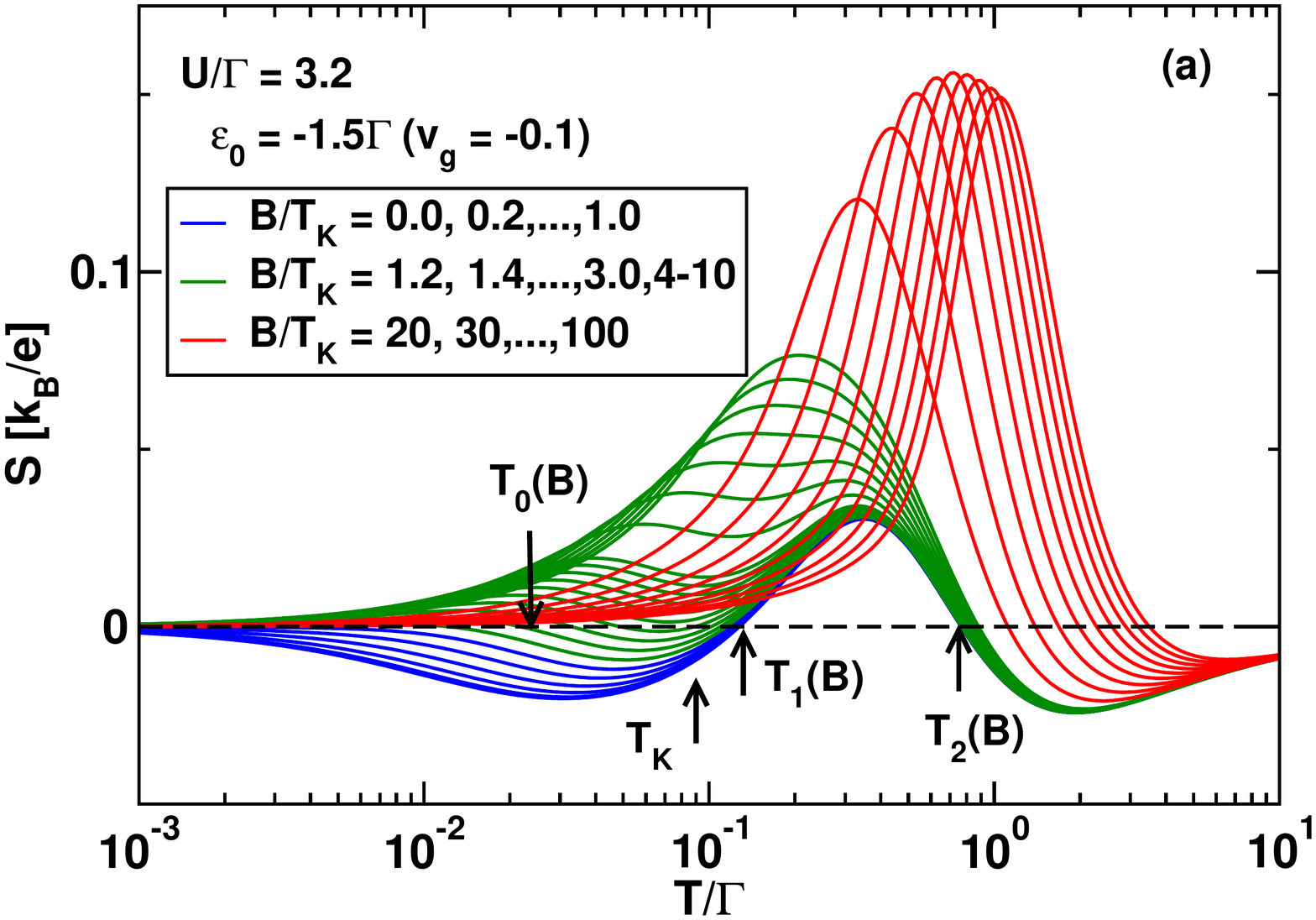}\hspace{1cm}
\includegraphics[width=0.95\columnwidth]{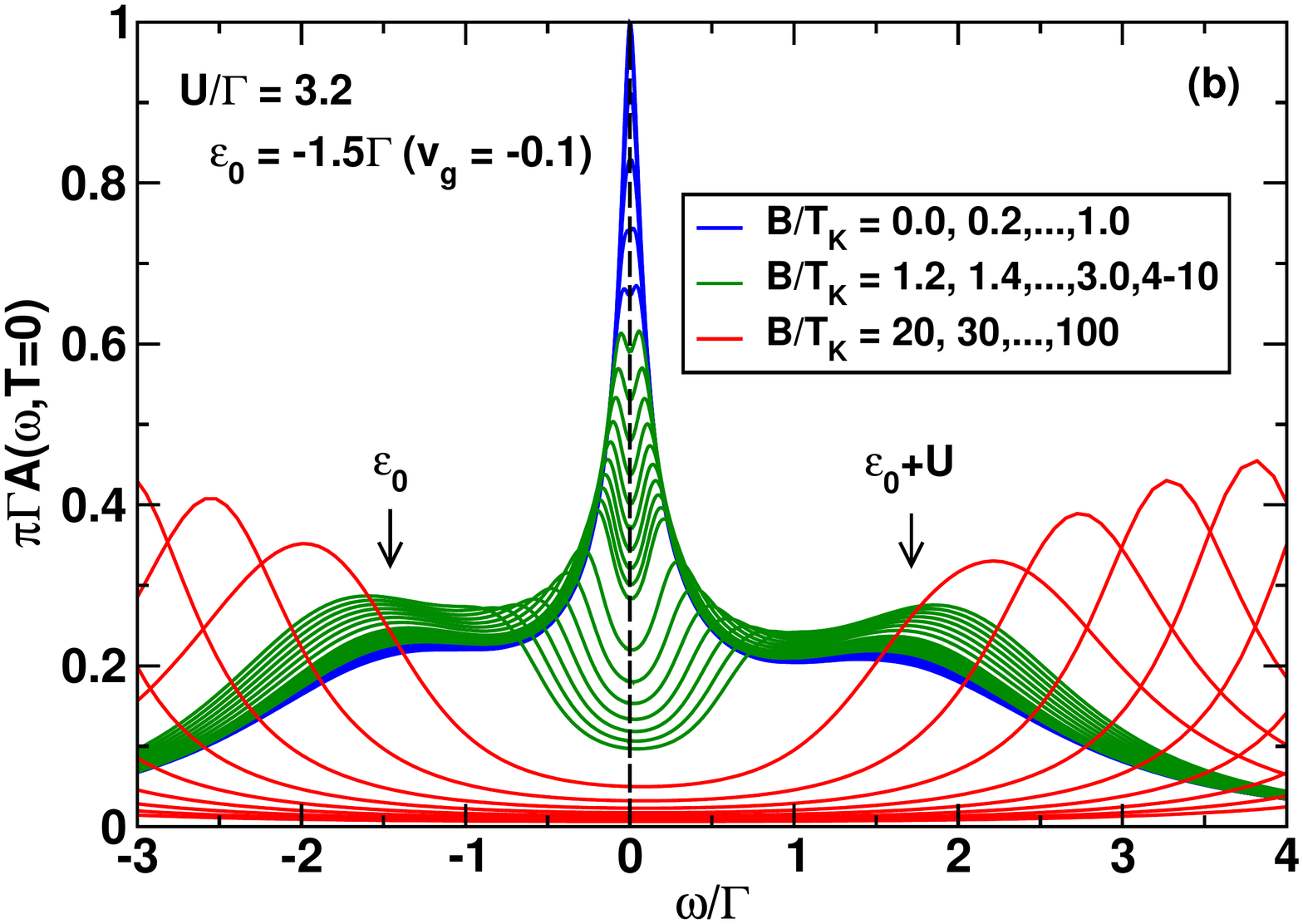}
\caption 
{(a) Thermopower $S$ (in units of $k_{\rm B}/e=86.17{\rm \mu V/K}$) vs temperature $T/\Gamma$, and, (b), $T=0$ total spectral function $A(\omega,T=0)$, for of a quantum dot with $U/\Gamma=3.2$, as in the experiment \cite{Svilans2018}, and for increasing values of the magnetic field $B/T_{\rm K}$ in the Kondo regime [$\varepsilon_{0}=-3\Gamma$ (${\rm v}_{g}=-0.1$), $T_{\rm K}/\Gamma = 9.192\times 10^{-2}$]. 
 For $B<B_{0}\approx 1.04T_{\rm K}$ (blue solid lines), two sign changes are found in $S(T)$ at $T_{1}(B)$ and  $T_{2}(B)$ , whereas for $B_{0}<B<B_{1}$ (green solid lines) an additional sign change occurs at a temperature $T_{0}(B)$, and, for $B>B_{1}$ (red solid lines), only the sign change at $T_{2}(B)$ is present in $S(T)$. In (b), $\varepsilon_0$ and $\varepsilon_0+U$ denote the bare Hubbard satellite excitations of the dot. NRG parameters: discretization parameter $\Lambda=4$, $z$ averaging \cite{Oliveira1994,Campo2005} with $N_z=4$, retaining $N_{\rm states}=900$ states. 
}
\label{fig:fig1}
\end{figure*}

\section{Kondo-induced sign changes in the thermopower}
\label{sec:sign-changes}
In this section, we describe the Kondo-induced sign changes in the thermopower $S(T)$ in the presence of a magnetic field for a quantum dot with $U/\Gamma=3.2$ (the value for the quantum dot QD1a in Ref.~\onlinecite{Svilans2018}, see Sec.~\ref{sec:comparison} for further details) and contrast these with the field-dependent behavior of the
thermopower $S(T)$ in the mixed valence and empty orbital regimes.

Figure~\ref{fig:fig1}(a) shows the temperature dependence of the thermopower $S(T)$ for a fixed gate voltage ${\rm v}_g=-0.1$ in the Kondo regime and increasing magnetic fields, while Fig.~\ref{fig:fig1}(b) shows the corresponding $T=0$ spectral functions of the dot.
For $B=0$ \cite{Costi2010}, $S(T)$ exhibits a (negative) Kondo-induced thermopower peak at $T\approx T_{\rm K}$ and two sign changes at the gate-voltage-dependent temperatures $T_{1}\gtrsim T_{\rm K}$ and $T_{2}\gtrsim \Gamma$, which are characteristic of the Kondo regime and are absent in the other regimes, where $S(T)$ is of one sign [see $B=0$ curves of Fig.~\ref{fig:fig2}, Figs.~\ref{fig:fig10}(a) and \ref{fig:fig11}(a) in Appendix~\ref{sec-mixed-valence} and Ref.~\onlinecite{Costi2010}]. While the physical significance of $T_{\rm K}$ as a low-energy scale of the Anderson model in the Kondo regime is clear \cite{Hewson1997}, that of $T_{1}$ or $T_2$ is more subtle.  Unlike $T_{\rm K}$, neither $T_{1}$ nor $T_2$ are low-energy scales since they are not exponentially small in $U/\Gamma$ \cite{Costi2010,Dutta2019}. Despite this, they are nevertheless closely connected to Kondo physics \cite{Costi2010,Dutta2019}. For example, the sign change at $T_1$ results from a subtle rearrangement of spectral weight in the asymmetrically located Kondo resonance within a region $-k_{\rm B}T \lesssim \omega \lesssim+k_{\rm B}T$ with increasing temperature \cite{Dutta2019}, while that at $T_2\gtrsim \Gamma$ is associated with a rearrangement of spectral weight in the high energy Hubbard satellite peaks at $\omega =\varepsilon_0$ and $\varepsilon_0+U$, whose weights are approximately given by $2-n_0(T)$ and $n_0(T)$, respectively, within the atomic limit approximation $t\to 0$ for the Anderson model.  Indeed, one finds for all level positions in the Kondo regime, that the value of $T_2$ correlates with a minimum (maximum) in $n_0(T)$ vs $T$ for ${\rm v}_g<0$ (${\rm v}_g>0$) corresponding to a significant spectral weight rearrangement at high energies \cite{Dutta2019}. Since the thermopower at temperatures $T=T_2 \gtrsim \Gamma$ probes the tails of the above excitations, a relative change in their weight can lead to the sign change observed at $T_2$.
We note that such a sign change, associated with a minimum (maximum) in $n_0(T)$ vs $T$, is only present in the Kondo regime \cite{Costi2010}. 
\begin{figure}[t]
  \centering 
  \includegraphics[width=0.95\columnwidth]{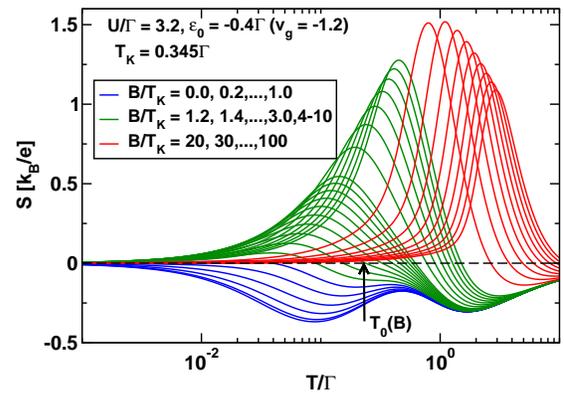}
  \caption 
  {Thermopower $S$  (in units of $k_{\rm B}/e=86.17{\rm \mu V/K}$) vs 
    temperature $T/\Gamma$ of a strongly correlated quantum dot ($U=3.2\Gamma$) for 
    increasing values of the magnetic field $B/T_{\rm K}$ and for local 
    level position $\varepsilon_0=-0.4\Gamma$ in the mixed valence 
    regime (${\rm v}_g=-1.2$).  Spin 
    susceptibility $T_{\rm K}\approx 0.345\Gamma$, is close to the mixed valence low-energy scale 
    $\Delta=0.5\Gamma$. 
    For $B<B_0\approx T_{\rm K}$, the thermopower 
    exhibits no sign change as a function of $T$ (blue solid lines), while for $B\gtrsim B_0$
    a single sign change at $T_{0}(B)$ occurs. NRG parameters as in Fig.~\ref{fig:fig1}. 
  }
  \label{fig:fig2}
\end{figure}
\begin{figure}[t]
\centering 
\includegraphics[width=0.95\columnwidth]{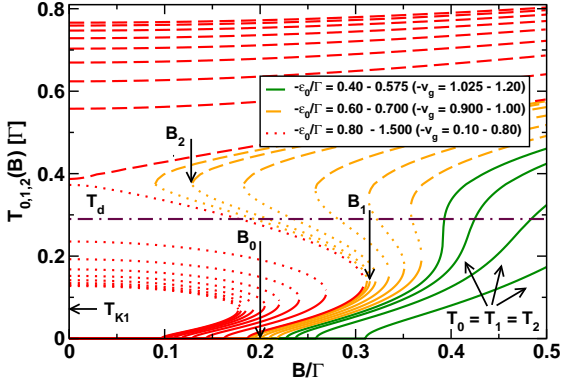}
\caption  
{$T_0(B)$ (solid lines), $T_1(B)$ (dotted lines) and $T_2(B)$ (dashed lines) in units of  
  $\Gamma$ vs $B/\Gamma$ for different $\varepsilon_0/\Gamma$ (${\rm v}_g$). 
  Outside the Kondo 
  regime ($|{\rm v}_g|\gtrsim 1.0$) only the sign change in $S(T)$ at $T=T_0(B)$
  exists. $T_d$ (dot-dashed line) is the highest temperature ($4.0\,{\rm K}$) in the experiment 
  \cite{Svilans2018} and $T_{\rm K1}$ indicates the midvalley (${\rm v}_g=0$) Kondo scale. 
  In detail, the local level positions $\varepsilon_0/\Gamma$ are as follows. Mixed valence regime (green lines): $-0.4,-0.5,-0.55$, and, $-0.575$. Weak Kondo regime (orange lines): $-0.6,-0.625,-0.65,-0.675,-0.6875$, and, $-0.69375$.  Kondo regime (red lines): $-0.7,-0.8,-0.9,-1.0,-1.1,-1.2,-1.3,-1.4$, and, $-1.5$. 
  NRG parameters as in Fig.~\ref{fig:fig1}. 
  }
\label{fig:fig3}
\end{figure}

At finite fields, the thermopower evolves as follows: for $B\lesssim T_{\rm K}$, the
thermopower $S(T)$ has a similar temperature dependence as for $B=0$, with two sign changes at $T_1(B)$ and $T_2(B)$,
where $T_{1}(B)$ and $T_{2}(B)$ are the finite-$B$ analogs of the two temperatures $T_1$ and $T_2$ where $S(T)$ changes sign at $B=0$. The main effect of $B$ on $S(T)$ in this low-field limit is to shift the  Kondo-induced peak in $S(T)$ at $T\approx T_{\rm K}$ to
higher temperatures and to reduce it in amplitude with increasing $B$, while leaving its sign unchanged [see Fig.~\ref{fig:fig1}(a)].  Once $B$ exceeds a gate-voltage-dependent value, $B_{0}$, but still below another field $B_1$ (to be discussed below), the thermopower exhibits an additional sign change at a temperature
$T_{0}(B)<T_{1}(B)<T_{2}(B)$. The meaning of $B_0$ follows from a Sommerfeld expansion for $S(T\to 0)$,
\begin{align}
  S(T) &\approx -\frac{k_{\rm B}}{|e|}\frac{\pi^2}{3}k_{\rm B}T \frac{1}{A(0,0)}\frac{\partial}{\partial\omega} A(\omega,T=0)|_{\omega=0},\label{eq:sommerfeld}
\end{align}
i.e., the sign change in $S(T)$  for $B>B_0$ reflects a change in sign of the slope 
of the $T=0$ spectral function at the Fermi level upon increasing $B$ through $B_0$. In the Kondo regime, the change in slope of the spectral function is brought about by a redistribution of spectral weight about the Fermi level as the asymmetrically located Kondo resonance splits with increasing magnetic field [Fig.~\ref{fig:fig1}(b)] and occurs on a comparable, but slightly larger field scale than that, $B_c$, for the splitting of the Kondo resonance, i.e., $B_{0}\gtrsim B_c$, as discussed in detail elsewhere \cite{Costi2019a}. This redistribution of spectral weight is highly nontrivial for the many-body Kondo resonance which remains pinned close to, but just above, the Fermi level with increasing low magnetic field [see Fig.~\ref{fig:fig1}(b)], so a discernible change in slope is barely visible.
In contrast, for the noninteracting resonance in the empty orbital regime at $\omega\approx \varepsilon_0>0$ [see  Fig.~\ref{fig:fig11}(b)] such a change in slope at the Fermi level for $B>B_0$ is a trivial effect of up-spin and down-spin components of the resonance moving in opposite directions and is clearly visible. In the mixed valence case, the low-energy resonance is renormalized by interactions to lie just above the Fermi level for $B=0$ [e.g., see Fig.~\ref{fig:fig10}(b)]]. With increasing magnetic field, this resonance splits into its up-spin and down-spin components, which move in opposite directions, resulting, for sufficiently large $B$, in a change in slope of the spectral function at the Fermi level [see Fig.~\ref{fig:fig10}(b)]. Hence, while the mixed valence and empty orbital regimes do not exhibit the sign changes in $S(T)$ at $T_1(B)$ and $T_2(B)$,
characteristic of the Kondo regime, they do exhibit a trivial sign change at $T_0(B)$ for $B>B_0$ [see Fig.~\ref{fig:fig2} and Figs.~\ref{fig:fig10}(a) and \ref{fig:fig11}(a) in Appendix~\ref{sec-mixed-valence}]. 

Further increasing $B$ towards a gate-voltage-dependent value $B_1$ results in a merging of $T_{0}(B)$ and $T_{1}(B)$ to a common value at $B=B_1$ [Fig.~\ref{fig:fig1}(a)] where $B_1$ is of order $T_1$ (see Appendix~\ref{sec:B1Gamma}). For $B>B_1$ (and for ${\rm v}_g$ still in the Kondo regime),  only the sign change at $T_2$ remains.

Thus, in the Kondo regime, a sign change in $S(T)$ at  $T=T_{0}(B)$ for $B_0<B<B_1$ is an additional characteristic feature of the Kondo effect in $S(T)$, in addition to the sign changes at $T_1(B)$ and $T_2(B)$. A further characteristic feature can be 
seen from Fig.~\ref{fig:fig1}(a), namely, for fixed ${\rm v}_g$ and fixed $T\lesssim T_1(0)$, the thermopower, $S_{T,{\rm v}_g}(B)$, has 
opposite signs for $B\to 0$ and $B>B_1$. This is also observed in the measurements of Ref.~\onlinecite{Svilans2018}, as discussed in Appendix~\ref{sec:GS-comparison}. Outside the Kondo regime,
only the field-driven sign change in $S(T)$ at $T=T_0(B)$ for $B>B_0$ is possible, which is seen to be a trivial one in this case (resulting from a trivial splitting of a weakly or noninteracting resonance in  a sufficiently large magnetic field).

The detailed evolution of $B_0$ and $B_1$ with gate voltage and different values of $U/\Gamma$ \cite{Costi2019a} has been described elsewhere \cite{Costi2019a} (see also Fig.~\ref{fig:fig7} in Sec.~\ref{sec:comparison}). We next turn to a description of the evolution of  $T_{0}(B),T_{1}(B)$ and $T_{2}(B)$ with $B$ for different gate voltages ranging from the Kondo to the mixed valence regime.  This is shown in Fig.~\ref{fig:fig3} for $U/\Gamma=3.2$. The same qualitative behavior of $T_{0}(B),T_{1}(B)$ and $T_{2}(B)$ vs $B$ is found for all $U/\Gamma\gg 1$, e.g., for $U/\Gamma=5$ \cite{Costi2019a}, while for weakly correlated quantum dots, e.g. for $U/\Gamma=1$, which do not exhibit the Kondo effect, the sign changes at $T_1$ and $T_2$ are absent both at $B=0$ \cite{Costi2010} and finite $B$. For all gate voltages, we note the general trends that $T_{0}(B)$ and $T_2(B)$ increase monotonically with increasing $B$, while $T_1(B)$ decreases monotonically with increasing $B$.
In the Kondo regime (red lines), we find a significant $B$-dependence in $T_{0}(B)$ and $T_{1}(B)$,  while $T_{2}(B)$ exhibits a weaker
dependence on field. Deep in the mixed valence regime (green lines), and also in the empty orbital regime, the single sign change at $T_0(B)$ is approximately linear in $B$ for $B\gg B_0$. The region between the mixed valence and the Kondo regime, which we labeled the ``weak Kondo regime'' (orange lines), exhibits features of both the mixed valence regime [absence of sign changes at $T_1(B)$ and $T_2(B)$ for $B<B_2$, where
$B_2$ depends on the gate voltage] and the Kondo regime [presence of sign changes at $T_1(B)$ and $T_2(B)$ but only for  $B>B_2>0$]. In this region, all $T_i(B),i=0,1,2$ exhibit a strong $B$ dependence. We note that the range of gate voltages corresponding to the weak Kondo regime is very narrow: $0.9\leq {\rm v}_g\leq 1.0$ ($-0.7 \leq \varepsilon_0/\Gamma \leq -0.6$). On approaching the mixed valence regime from the weak Kondo regime, we see that the temperatures $T_0(B), T_1(B)$ and $T_2(B)$ merge to the single temperature $T_0(B)$, with $T_0(B)$ exhibiting an inflexion point when $B_2=B_1$.
We also note, in connection with the experiment \cite{Svilans2018}, that $T_2(B)$ lies above the highest temperature ($T_d=4.0 \,{\rm K}$) of the experiment, so in comparing with experiment in Sec.~\ref{sec:comparison} we need only consider the sign changes at $T_{0}(B)$ and $T_{1}(B)$.
\section{Gate-voltage dependence of the thermopower}
\label{sec:gate-dependence}
\begin{figure}[t]
\centering 
  \includegraphics[width=0.95\columnwidth]{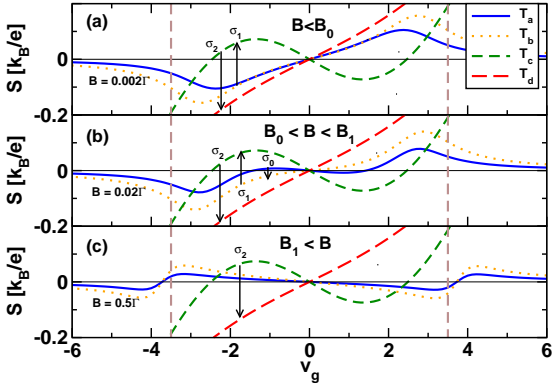}
  \caption{
     $S$ vs ${\rm v}_g$ at four temperatures $T_a<T_b<T_c<T_d$ for, (a), $B<B_0$, (b), $B_0<B<B_1$, and, (c), $B>B_1$, showing the possible sign 
    changes (labeled $\sigma_0,\sigma_1$ and $\sigma_2$) of $S$ in the Kondo regime upon increasing temperature through $T_{0}, T_{1}$ and $T_{2}$ (for given fixed ${\rm v}_g$ in the Kondo regime) . $U=8\Gamma$ and 
    $B_0({\rm v}_g\to 0)\approx 0.004\Gamma$, $B_1({\rm v}_g\to 0)\approx 0.05\Gamma$. 
     Specifically, the sign changes involved are, (a), $\sigma_1$: $T_{a,b}<T_1(B)<T_c$ and $\sigma_2$: $T_c<T_2(B)<T_d$, (b), as in (a) and the additional sign change $\sigma_0$: $T_a<T_0(B)<T_b$, (c), only the sign change $\sigma_2$: $T_c<T_2(B)<T_d$.  
    Temperatures   $T_{a,b,c,d}/\Gamma=0.0038,0.0084,2.222$ and $3.62$ chosen relative to 
    midvalley estimates $T_1/\Gamma=0.00426$ and $T_2/\Gamma=2.69$ to observe 
    the above sign changes. 
    Vertical dashed lines 
    delineate Kondo ($|{\rm v}_g|\lesssim 3.5$) from mixed valence and empty (full) orbital regimes at larger $|{\rm v}_g|$. 
    NRG parameters as in Fig.~\ref{fig:fig1}. 
  }
\label{fig:fig4}
\end{figure}
 Experiments on quantum dots probe the thermoelectric response as a function of gate voltage at fixed temperature and fixed magnetic field. Hence, in this section we show how the Kondo-induced sign changes in the thermopower are reflected in $S$ vs ${\rm v}_g$ at fixed $B$ and $T$. From the previous section, we see that three  field ranges determine the possible sign changes: (a) $B<B_0$, (b), $B_0<B<B_1$, and, (c), $B>B_1$. Since the fields $B_0$ and $B_1$ also depend on ${\rm v}_g$, we shall
here discuss the generic behavior expected in the Kondo regime close to midvalley (${\rm v}_g=0$). Thus,
for case (a) we expect two sign changes in $S$ at $T_1(B)$ and $T_2(B)$ upon increasing $T$ at fixed ${\rm v}_g$, for case (b) we expect in addition a sign change at $T_0(B)$, and for case (c) we expect only
the sign change at $T_2(B)$. To illustrate these cases, we choose $U/\Gamma=8$. Using midvalley values
for $B_0$ and $B_1$ we choose appropriate fields $B$ for each case, and appropriate temperatures $T_a<T_b<T_c<T_d$ to manifest the sign changes at $T_0(B),T_1(B)$ and $T_2(B)$. This is shown in
Fig.~\ref{fig:fig4}.  In case (a), the chosen temperatures satisfy $T_a<T_b<T_1<T_c<T_2<T_d$ and sign changes at $T_1$ (denoted by $\sigma_1$) and at $T_2$ (denoted by $\sigma_2$) are found, as expected. In case (b), the chosen temperatures now satisfy 
$T_a<T_0<T_b<T_1<T_c<T_2<T_d$ and exhibit the additional sign change at $T_0(B)$ (denoted by $\sigma_0$). Finally, for case (c), the chosen temperatures satisfy  $T_a<T_b<T_c<T_2<T_d$ and, as expected, for $B>B_1$, only the sign change on increasing temperature through $T_2(B)$ is observed.
\begin{figure}[t]
\centering 
\includegraphics[width=0.95\columnwidth]{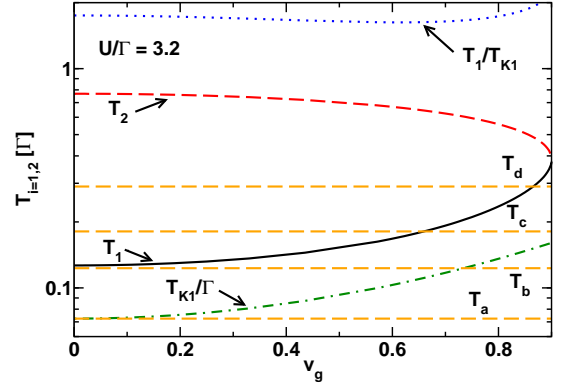}
\caption  
{$T_1$ and $T_2$ vs ${\rm v}_g$ in the Kondo regime for $U/\Gamma=3.2$ and 
  $B=0$. Also indicated is $T_1/T_{\rm K1}$ vs ${\rm 
    v}_g$ and $T_{\rm K1}$ vs ${\rm 
    v}_g$, where $T_{\rm K1}$ is the perturbative Kondo scale Eq.~(\ref{eq:tk-vs-vg}) as used 
  in the experiment \cite{Svilans2018}. The dot-dashed horizontal 
  lines indicate the four experimental temperatures $T_{a,b,c,d}/\Gamma=0.0725,0.123,0.181$, 
  and $0.290$. NRG parameters as in Fig.~\ref{fig:fig1}. 
}
\label{fig:fig5}
\end{figure}

An alternative quantity that probes the Kondo-induced sign changes in the thermopower is the slope
of the linear-response thermocurrent $I_{\rm th}/\Delta T = G(T)S(T)$ with respect to ${\rm v}_g$ at ${\rm v}_g=0$, i.e., $\sigma(T)=d [G(T)S(T)]/d{\rm v}_{g}|_{{\rm v}_g=0}$ (for definitions, see next section).
Clearly, this exhibits exactly the same sign changes at $T_0(B), T_1(B)$ and $T_2(B)$ as $S(T)$
close to midvalley. It has been compared with relevant measurements \cite{Svilans2018}
in Ref.~\onlinecite{Costi2019a}, so we do not discuss this further here.

\section{Comparison with experiment}
\label{sec:comparison}
We now compare our results to measurements of the thermoelectric response of {\rm InAS} quantum dots \cite{Svilans2018}. We first note, that the linear-response current, $I_{\rm SD}$, through a quantum dot subject to a temperature difference $\Delta T$ and a bias voltage $\Delta V_{\rm bias}$ across the leads is given by \cite{Prete2019} 
\begin{align}
  I_{\rm SD} = G(T) \Delta V_{\rm bias} + G(T) S(T) \Delta T,\label{eq:linear-response}
  \end{align}
  where $T$ is the average temperature of the two leads and $G(T)$ is the linear conductance at temperature $T$. The induced voltage under open circuit conditions ($I_{\rm SD}=0$) is the thermovoltage $\Delta V_{\rm th}\equiv \Delta V_{\rm bias} ^{I_{\rm SD}=0}$ which, from Eq.~(\ref{eq:linear-response}), yields the thermopower $S(T)=-\Delta V_{\rm th}/\Delta T$ studied in this
  paper. A different measure of the thermoelectric response has been investigated  in Ref.~\onlinecite{Svilans2018}, namely the current resulting from a temperature gradient $\Delta T$ at zero bias, i.e., the thermocurrent $I_{\rm th}=I_{\rm SD}^{\Delta V_{\rm bias}=0}$. From Eq.~(\ref{eq:linear-response}), we have that $I_{\rm th}/\Delta T = G(T) S(T)$, i.e., the thermocurrent
  measured in Ref.~\onlinecite{Svilans2018} is proportional, within linear-response, to $S(T)$, up to a temperature dependent prefactor $G(T)$. In the following, we work on the assumption that the measurements for $I_{\rm th}/\Delta T$ were in the linear-response
    regime, and compare these with our linear-response calculation for the same quantity $G(T)S(T)$\footnote{
For the differences between $S$ and $GS$, see Appendix~\ref{sec:GS-comparison}.}
    . Under the same assumption,
    it is clear that the measured thermocurrent exhibits the same sign changes at $T_{i=0,1,2}(B)$ as those in $S(T)$. We return
    to this, and other assumptions, in Sec.~\ref{sec:assumptions}.

  We focus on device QD1a of Ref.~\onlinecite{Svilans2018}, which has $U=3.5\, {\rm meV}$, and $\Gamma=1.1\, {\rm meV}$ ($U/\Gamma=3.2$), resulting in a midvalley 
$T_{\rm K1}^{\rm exp}\approx 1.0\, K$. With these parameters, we find that $T_{1}/T_{\rm K1}$ is a weak function of gate voltage in the Kondo regime, with $1.62 \lesssim T_{1}(B=0)/T_{\rm K1}\lesssim 1.75$ (Fig.~\ref{fig:fig5}) , consistent with the experimentally cited value of $1.8$ at midvalley \cite{Svilans2018}. Similarly, for $T_2(0)$, we find that  $5.0\, K \lesssim T_{2}(0)\lesssim\, 11.0\, K$ (corresponding to $0.38 \lesssim T_{2}(0)/\Gamma \lesssim 0.77$, see Fig.~\ref{fig:fig5}),
i.e., the sign change in the thermopower at $T_2(0)$ occurs above the highest temperature ($4.0\,K$)  of the experiment and therefore need not be considered further. From the value of $\Gamma$, and the measured $g$ factor $g\approx 9$ for {\rm InAs} quantum dots \cite{Kretinin2011,Svilans2018}, we carry out calculations for $GS$ vs ${\rm v}_g$
at the experimental field values ($B_{a,b,c,d}=0.0\,T ,0.5\,T, 1.0\,T$ and $2.0\,T$) and temperatures ($T_{a,b,c,d}=1.0\,K, 1.7\,K, 2.5\,K$ and $4.0\,K$) \cite{Svilans2018}. The four field values correspond to $B_{a,b,c,d}/T_{\rm K1}(0)\approx 0,3,6$ and $12$. For convenience, Table~\ref{Table1} lists the values of these temperatures and fields in physical units, and also in
units of $\Gamma$ as used in the model calculations (see also last sentence of Sec.~\ref{sec:model+transport}).
\begin{table}[t]
\begin{ruledtabular}
\begin{tabular}{cccccc}
\multicolumn{1}{c}{$i$} & 
\multicolumn{1}{c}{$T_{i}[{\rm K}]$}&
\multicolumn{1}{c}{$T_{i}[\Gamma]$}& 
\multicolumn{1}{c}{$B_{i}[{\rm T}]$}& 
\multicolumn{1}{c}{$B_{i}[\Gamma]$}& 
\multicolumn{1}{c}{$B_{i}[T_{\rm K1}(0)]$}\\
\colrule 
$a$   & $1.0$ & $0.0725$            & $0.0$  & $0.0$ & $0$\\ 
$b$  & $ 1.7$ &$0.123$             & $0.5$  & $0.21735$& $3$\\ 
$c$   & $ 2.5$ &$0.181$             & $1.0$  & $0.4347$& $6$\\ 
$d$   & $ 4.0$ &$0.290$             & $2.0$  & $0.8694$& $12$\\ 
\hline 
\end{tabular}
\end{ruledtabular}
\caption 
{ Temperatures $T_{i=a,b,c,d}$ and magnetic fields $B_{i=a,b,c,d}$ used in the experiment \cite{Svilans2018} in physical units and units of $\Gamma$ ($T_i/\Gamma$, $B_i/\Gamma$ to be read 
  as $k_{\rm B}T_i/\Gamma$, $g\mu_{\rm B}B_i/\Gamma$ respectively). Also listed is the approximate value for $B_i/T_{\rm K1}(0)$, where $T_{\rm K1}(0)$ is the midvalley Kondo scale for $U/\Gamma=3.2$. 
}
\label{Table1}
\end{table}
\begin{figure*}[t]
\centering 
\includegraphics[width=0.95\columnwidth]{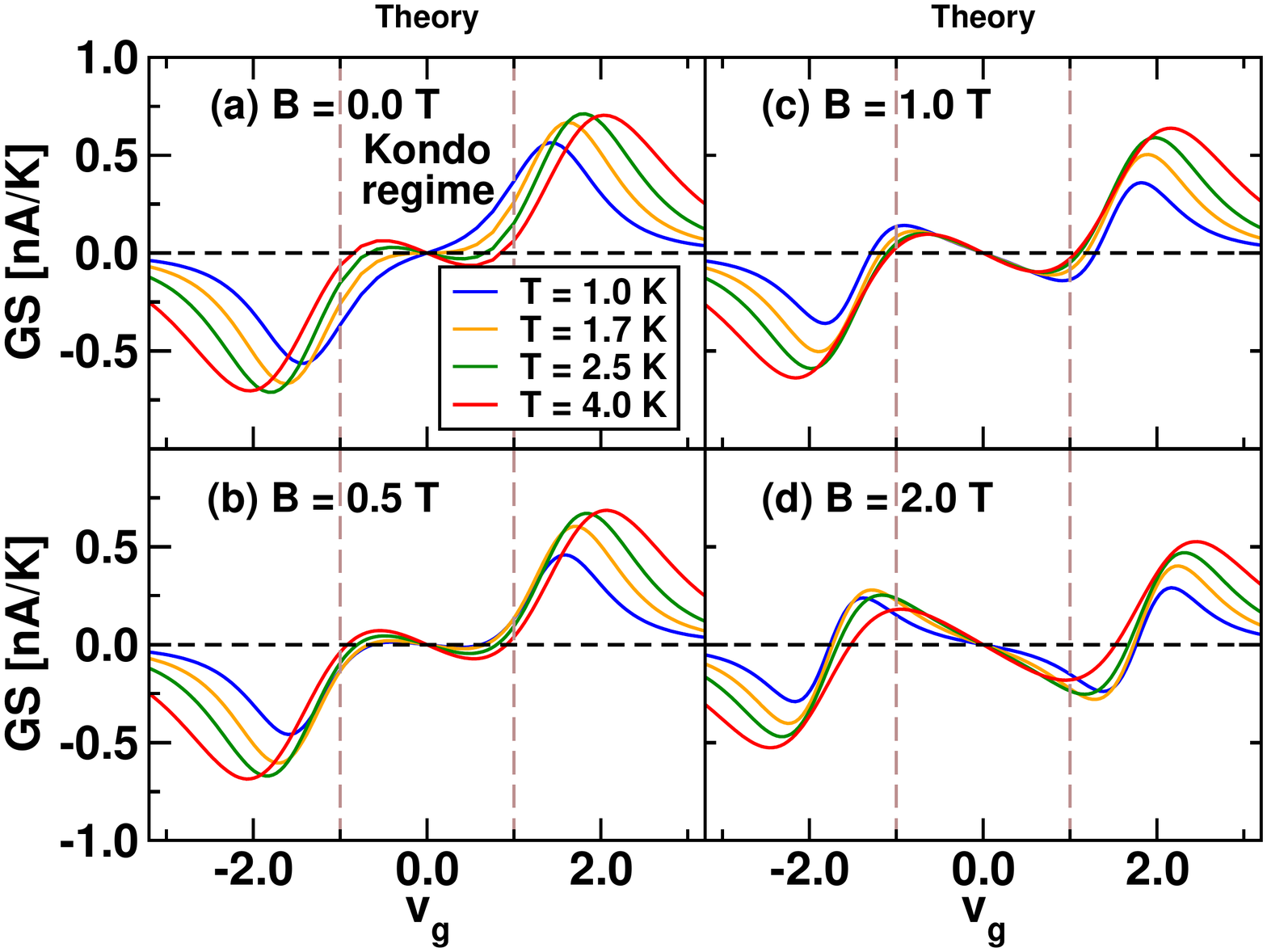}\hspace{1cm}
  \includegraphics[width=0.95\columnwidth]{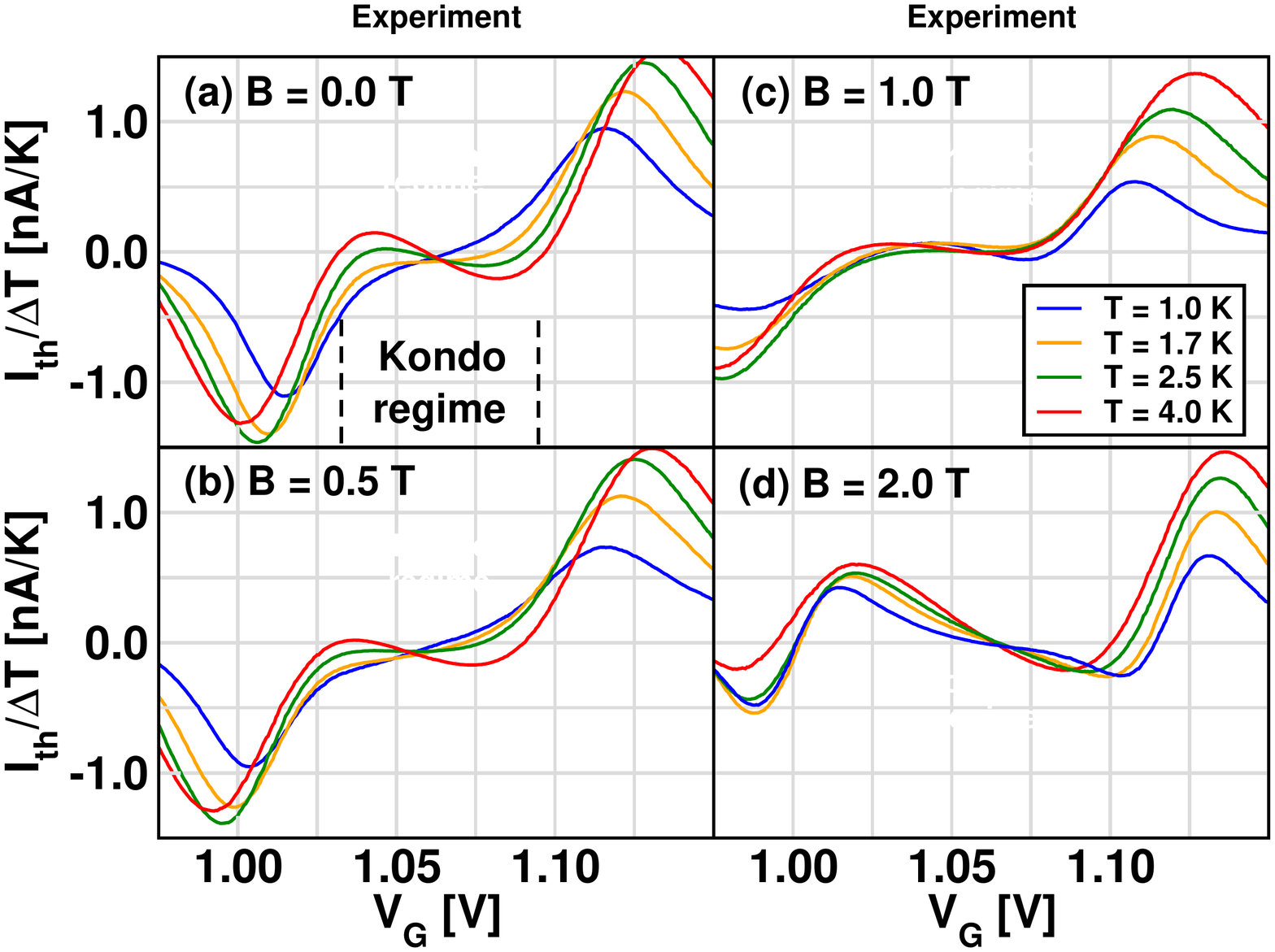}
  \caption{
Left panels (a)-(d): calculated $GS$ vs ${\rm v}_g$ at the temperatures $T$ and the four magnetic fields $B$ of the experiment \cite{Svilans2018}. 
Vertical dashed lines delineate Kondo ($|{\rm v}_g|\lesssim 1.0$) from mixed valence and empty (full) orbital regimes at larger $|{\rm v}_g|$. 
  Right panels (a)-(d): experimentally measured thermocurrent $I_{\rm th}/\Delta T$ vs gate-voltage $V_G$
  at four magnetic fields and temperatures for device QD1a (adapted, with permission, from Ref.~\onlinecite{Svilans2018}). 
  NRG parameters 
  as in Fig.~\ref{fig:fig1}.}
\label{fig:fig6}
\end{figure*}

The results are shown in Figs.~\ref{fig:fig6}(a)-\ref{fig:fig6}(d) (left four panels), and are to be compared with the corresponding experimental results from Ref.~\onlinecite{Svilans2018} shown in the right four panels. The resemblance of our results to those of the experiment is quite striking. Starting with some general observations, we note that outside the Kondo regime $|{\rm v}_g|\gtrsim 1.0$ [delineated by
vertical dashed lines in Figs.~\ref{fig:fig6}(a)-\ref{fig:fig6}(d)], much the same overall trends with increasing temperature are observed in both theory and experiment, e.g., the similar increase in magnitude of $GS$ with increasing temperature and the lack of a significant $B$ dependence for
$|{\rm v}_g|\gtrsim 1.0$.
More striking, are the strong similarities between theory and experiment in $GS$ vs ${\rm v}_g$ in the Kondo regime of gate-voltages, at each $B$, and for increasing temperature: for example, the significant temperature variation of $GS$ at the lowest fields, compared to the near absence of a temperature variation in the case of $B=1.0\,T$ and the recovery of some temperature variation at $B=2.0\,T$. For the latter case, note, in particular, the inflection of the $T=1.0\,K$ curve at midvalley, present in both theory and experiment. Thus, for all four field values the overall temperature trends in $GS$ vs ${\rm v}_g$ are strikingly similar between theory and experiment and the order of magnitude of the response (up to $\approx 0.75\,nA/K$ in theory, and up to $\approx 1.5\,nA/K$ in experiment) is the same for both.

\begin{figure}[t]
\centering 
\includegraphics[width=0.95\columnwidth]{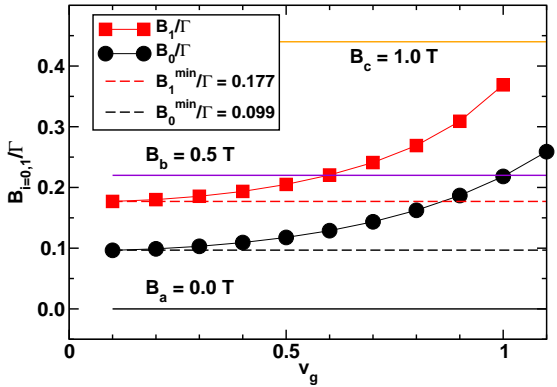}
\caption  
{$B_{0}$ and $B_{1}$ (in units of $\Gamma$) vs ${\rm v}_g$ for 
  $U/\Gamma=3.2$. Also indicated (in units of $\Gamma$) 
  are the minimum values of $B_{0}$ ($B_{0}^{\rm min}$) and $B_{1}$
  ($B_{1}^{\rm min}$) at ${\rm v}_g=0$
  (dashed horizontal lines) and the three lowest 
  experimental fields $B_{a,b,c}=0.0\,T, 0.5\,T$ and $1.0\,T$ in 
  Ref.~\onlinecite{Svilans2018} (solid horizontal lines). The highest 
  experimental field, $B_{d}=2.0\,T$, lies much above $B_{0}$ and 
  $B_{1}$ for the indicated range of ${\rm v}_g$ and is not shown. 
  NRG parameters as in Fig.~\ref{fig:fig1}.
}
\label{fig:fig7}
\end{figure} 
\begin{figure}[t]
\centering 
\includegraphics[width=0.95\columnwidth]{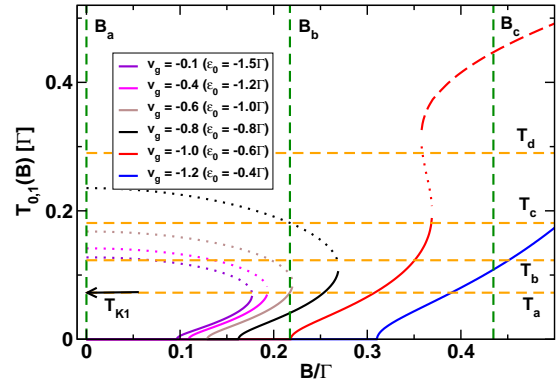}
\caption  
{$T_0(B)$ (solid lines) and $T_1(B)$ (dotted lines)  in units of  
  $\Gamma$ vs $B/\Gamma$ for different ${\rm v}_g$ [$T_2(B)>T_d$ and is not shown, being off the scale of the plot,
  (except for ${\rm v}_g=-1.0$)].
  Outside the Kondo regime (${\rm v}_g\gtrsim 1.0$) only the sign change in the 
  thermopower at $T_0(B)$  exists. The horizontal  
  dashed lines correspond to the four experimental temperatures $T_{a,b,c,d} =
  1.0\, K, 1.7\, K, 2.5\, K$ and $4.0\, K$
    , and the vertical dashed lines correspond to the 
    experimental fields $B_{a,b,c,d}=0.0\, T, 0.5\, T, 1.0\, T$ in
    Ref.~\onlinecite{Svilans2018}  (the highest field  $B=2.0\, T$
    is outside the scale of the plot). The midvalley Kondo scale $T_{\rm K1}$ (horizontal
    arrow) lies very close to the lowest experimental temperature $T_a$.
    NRG parameters as in Fig.~\ref{fig:fig1}.
  }
\label{fig:fig8}
\end{figure} 
The detailed $B$-dependence of $GS$ in
Figs.~\ref{fig:fig6}(a)-\ref{fig:fig6}(d) and of $I_{\rm th}/\Delta T$ in the measurements in Ref.~\onlinecite{Svilans2018} can be understood  depending on whether, (i), $B<B_0$, (ii), $B_0< B < B_1$, or, (iii),  $B>B_1$. For each field, we can determine which case applies by referring to Fig.~\ref{fig:fig7} which shows the ${\rm v}_g$-dependence of $B_0$ and $B_1$ relative to the fields $B_{i=a,b,c,d}$ in the experiment, while the
temperatures $T_0(B)$ and $T_1(B)$, at which sign changes in $G(T)S(T)$ at fixed ${\rm v}_g$ upon increasing $T$ can occur can be determined from Fig.~\ref{fig:fig8}, which shows the field and ${\rm v}_g$-dependence of $T_0$ and $T_1$. In comparing the gate-voltage dependence of $GS$ with the experimental thermoelectric response we shall focus on ${\rm v}_g<0$. In this context, it is useful to note that $B_0,B_1$ as well as $T_0,T_1$ and $T_2$ are symmetric functions of gate-voltage.
\subsection{$B=0.0\, {\rm T} < B_{0}$}
Starting with $B=0<B_0( {\rm v}_g)$ in Fig.~\ref{fig:fig6}(a), one confirms the zero field sign change in $G(T)S(T)$ at $T=T_1$ in the Kondo regime $|{\rm v}_g|\lesssim 1.0$, i.e., $G(T)S(T)$ at the two lowest temperatures $T=1.0\,K$ and $1.7\,K$  has an opposite sign to that at the two highest temperatures $T=2.5\,K$ and $4.0\,K$, as seen also in experiment (and consistent with the former being at $T<T_1$ and the latter at $T>T_1$, Fig.~\ref{fig:fig5}).

\subsection{$B=0.5\, {\rm T}$}
\begin{figure}[t]
  \centering 
  \includegraphics[width=0.95\columnwidth]{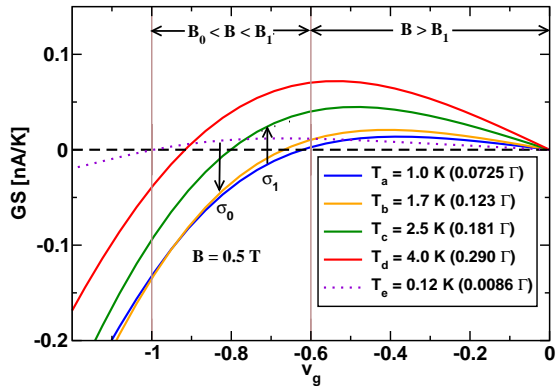}
  \caption  
  {Detailed view of  Fig.~\ref{fig:fig6}(b) for the calculated $GS$ vs ${\rm v}_g$ for ${\rm v}_g<0$ at $B=0.5\,T$ and at each 
    temperature $T_{a,b,c,d}$ of the experiment, and at one additional lower temperature $T_e=0.12\,{\rm K}$. Regions of gate voltage satisfying $B_{0}({\rm v}_g)<B<B_1({\rm v}_g)$ and $B_{1}({\rm v}_g)<B$ are indicated. While the sign change at $T_1$ (denoted $\sigma_1$) is observable for the experimentally used temperatures, e.g., on increasing temperature from $T_a$ to $T_c$, the sign change at $T_0(B)$ (denoted $\sigma_0$) in the range $-1.0\lesssim {\rm v}_g\lesssim -0.6$,
    is only observable by starting from the lower temperature $T_e <T_0(B)$, e.g., upon increasing temperature from $T_e$ to $T_a$. 
    NRG parameters as in Fig.~\ref{fig:fig1}.
  }
  \label{fig:fig9}
\end{figure}

For $B=0.5\,T$, we find  that $B > B_1( {\rm v}_g)$ for $-0.6\lesssim {\rm v}_g\lesssim 0.0$ (see Fig.~\ref{fig:fig7}), so $GS>0$, as seen in both theory and experiment. For a small range, $-1.0\lesssim {\rm v}_g\lesssim -0.6$, we find that $B_0( {\rm v}_g)<B = 0.5\,{\rm T}<B_1( {\rm v}_g)$ , so $GS$ could, in principle, show the sign change at $T_0(B)$ (from $GS>0$  to $GS<0$) upon increasing $T$ through $T_{0}(B)$, in addition to the one at $T_1>T_0$ (from $GS<0$ to $GS>0$). However, since the four temperatures of the experiment all lie above $T_0(B)$, this sign change is not observed in the experiment (in contrast to the sign change at $T_1$, which is observed in the experiment, e.g., between $T=2.5\, K$ and $4.0\, K$ in Ref.~\onlinecite{Svilans2018}). We elucidate this further by estimating
$T_0(B)$ at the ends of the interval $-1.0 \lesssim {\rm v}_g \lesssim -0.6$.  
According to  Fig.~\ref{fig:fig8},
$T_0(B=0.5\, T) = 0.073$ (i.e., $T_{0}(B)=1.008\, K$) for ${\rm v}_g=-0.6$
and $T_0(B=0.5\, T) \approx 0.0\Gamma$ (i.e., $T_0(B)=0.0\,K$) for ${\rm v}_g=-1.0$.
Since  both of these values lie at, or,
below the lowest temperature $T=0.0725\Gamma$ ($1 \,K$) of the 
experiment, the sign change at $T_0(B)$ was not observed.
In order to observe this sign change, consider the gate voltage ${\rm v}_g=-0.8$. From
Fig.~\ref{fig:fig8}, for $B=0.5\,T$, we have
$T_0(B=0.5\, T) \approx 0.04\Gamma$ ($T_{0}(B)=0.55\, K$). Thus, a measurement of $GS$ at ${\rm v}_g=-0.8$ at a temperature below $0.55\,K$ will have $GS>0$, while a measurement at a
temperature above $0.55\,K$ will have $GS<0$, thereby showing the sign change at $T_0(B)$.
Figure~\ref{fig:fig9} shows the case $B=0.5\, {\rm T}$ 
from Fig.~\ref{fig:fig6}(b) in more detail for ${\rm v}_g<0$.
In addtion to the temperatures of the 
experiment $T_{a,b,c,d}>T_0(B)$, an additional, lower, temperature
$T_e= 0.086\Gamma$ ($T=0.12\, K$) is shown satisfying $T_e<T_0(B)$ (for gate-voltages
in the approximate range $-1.0 \lesssim {\rm v}_g \lesssim -0.6$). 
While, the experimentally used temperatures suffice to measure the
sign change at $T_1$, e.g., from $T=1.0\,{\rm K}$ with $GS<0$ to $T=2.5\,{\rm K}$ with $GS>0$,
denoted by $\sigma_1$ in Fig.~\ref{fig:fig9}, the sign change at $T_0$, denoted by $\sigma_0$ in Fig.~\ref{fig:fig9}, requires measuring from a temperature $T_{e}=0.12\,{\rm K}<T_0(B)$ (with $GS>0$)
to a higher temperature, e.g., $T=1.0\,{\rm K}$ (with $GS<0$).

\subsection{$B=1.0\, {\rm T}$ and $2.0 {\rm T}$}
For the two largest fields, $B=1.0\, T$ and $2.0\, T$, we are in the case $B>B_1$ for all gate-voltages in the Kondo regime (Fig.~\ref{fig:fig7}), hence $GS>0$ for $-0.9 \lesssim {\rm v}_g\lesssim 0$ as observed in theory [Fig.~\ref{fig:fig6}(c) and \ref{fig:fig6}(d)] and experiment.

  \subsection{linear-response and model assumptions}
  \label{sec:assumptions}
While our calculations for the linear-response thermocurrent $G(T)S(T)$ explain many of the trends observed in the measured thermocurrent $I_{\rm th}/\Delta T$,  which follow those predicted for a Kondo-correlated quantum dot, perfect quantitative agreement cannot be expected for several reasons. First, linear-response is certainly expected to be quantitatively accurate for $\Delta T \ll T_{\rm K1}$, but this (stringent) condition is not met in experiment, where $\Delta T\approx 0.3 T$-$0.35 T$ for $T$  in the temperature range $1.0-4.0\, K$, so that $\Delta T$ can range from $0.3$ to $1.4 T_{\rm K1}$ for $T_{\rm K1}\approx 1.0\, K$\cite{Svilans2018}. Secondly,
it is challenging to obtain good estimates of temperature gradients in experiments on quantum dots, and this could impact on the
magnitude of $I_{\rm th}/\Delta T$. Finally, we are making the approximation that only a single level of the quantum dot contributes to the transport. This is expected to be a good approximation in the Kondo regime, but to deteriorate in the other regimes, when further levels enter the transport window and become relevant.

\section{Conclusions}
\label{sec:conclusions}
In summary, we extended our field-dependent study of the thermopower of Kondo-correlated quantum dots \cite{Costi2019a}
to the mixed valence and empty orbital regimes, and characterized the detailed evolution of the Kondo-induced signatures
in the thermopower, quantified by $T_0(B),T_{1}(B)$, and $T_2(B)$, as a function of magnetic field and gate-voltage.
On approaching the mixed valence regime, the above temperatures coalesce to a single temperature $T_0(B)$, which is finite
in all three regimes for $B>B_0$, where $B_0$ is a gate-voltage-dependent field of order $T_{\rm K}$ in the Kondo regime
and of order $\Gamma$ in the mixed valence and empty orbital regimes. In all cases, $B_0$ corresponds to the field at
which the low-energy resonance in the $T=0$ spectral function changes slope at the Fermi level and is comparable to, but larger than, the field $B_c$, where this resonance splits in a magnetic field. While the sign change in the slope of the spectral function for $B>B_0$ is expected in the mixed valence and empty orbital regimes due to the weakly or almost noninteracting nature of their low-energy resonances, such a sign change in the Kondo regime is nontrivial because the Kondo resonance is a many-body singlet resonance strongly pinned close to the Fermi level [see Fig.~\ref{fig:fig1}(b)]. Hence, accurate NRG calculations seem imperative in order to capture this effect quantitatively. As shown elsewhere \cite{Costi2019a},
higher-order Fermi-liquid calculations for the spectral function \cite{Oguri2018a} can also capture this effect.

In the Kondo regime, three cases apply for the sign changes in $S(T)$: (i) $B<B_0$, with sign changes at $T_{1}(B)$ and $T_2(B)$, (ii), $B_0<B<B_1$, with the additional sign change at $T_0(B)$, and, (iii), $B>B_1$, where only the sign change at $T_2(B)$ remains. By carrying out detailed calculations for the gate-voltage dependence of $B_0$ and $B_1$ and that of
$T_0(B),T_{1}(B)$, and $T_2(B)$, using experimental parameters, we were able to compare our results for the gate-voltage dependence of the linear-response thermocurrent $GS$ with the corresponding measurements for the thermoelecrtric response of a Kondo-correlated quantum dot in Ref.~\onlinecite{Svilans2018}. The overall trends in the measured gate-voltage dependence at the fields and temperatures of the experiment are well recovered. We also showed, that while the Kondo-induced sign change at $T_1(B)$ is indeed observed in this experiment, observation of the sign change at $T_0(B)$, which according to theory can be realized for
the $B=0.5\, {\rm T}$ data, would require a temperature $T\approx 0.12\,{\rm K}$ below the lowest temperature measured ($1.0\,{\rm K}$). It would also be interesting to test our predictions for $S$ vs ${\rm v}_g$ by a direct measurement of the thermovoltage
(and hence Seebeck coefficient $S$), see discussion following Eq.~(\ref{eq:linear-response}), and
Refs.~\cite{Dutta2019,Prete2019}.

In contrast to electrical conductance ($G(T)$) measurements which probe primarily the excitations at the Fermi level, so that $G(T)$ is roughly proportional to the height of the Kondo resonance $A(0,T)$, thermopower measurements probe the relative importance of electronlike and holelike excitations. They give additional information on the low-energy Kondo resonance, such as its position relative to the Fermi level and how the relative weight below and above the Fermi level changes with temperature and magnetic field as reflected in the sign changes discussed in this paper.
Beyond being of relevance to experiments which characterize the thermoelectric properties of nanodevices \cite{Cui2017,Josefsson2018,Svilans2018,Dutta2019,Prete2019}, calculations along the same lines, can be carried out for classical Kondo impurities, and could be of some relevance to thermopower measurements in heavy fermions. In this paper we addressed only the
linear-response thermopower (and thermocurrent). Nanoscale devices, however, can be routinely driven out of equilibrium \cite{DeFranceschi2002, Leturcq2005,Josefsson2018}, and studying their nonequilibrium charge and heat currents with appropriate theoretical techniques \cite{Hershfield1991,Hershfield1993,Meir1993,Koenig1996,Rosch2003a,Anders2008a,Mehta2008,Leijnse2010,Moca2011,Pletyukhov2012,Munoz2013,Haupt2013,Dorda2016,Fugger2018,Schwarz2018,Oguri2018a,Eckern2019} is an interesting topic for future research.
\begin{acknowledgments}
  We thank A. Svilans for sending us the experimental data from Ref.~\onlinecite{Svilans2018}, used in
 the comparisons shown in  Fig.~\ref{fig:fig6}.
  Supercomputer support by the John von Neumann institute for Computing (J\"ulich) is also acknowledged.
\end{acknowledgments}
\appendix
\section{Magnetic field dependence of the thermopower in the mixed valence and empty orbital regimes}
\label{sec-mixed-valence}
\begin{figure*}[t]
  \centering 
  \includegraphics[width=0.95\columnwidth]{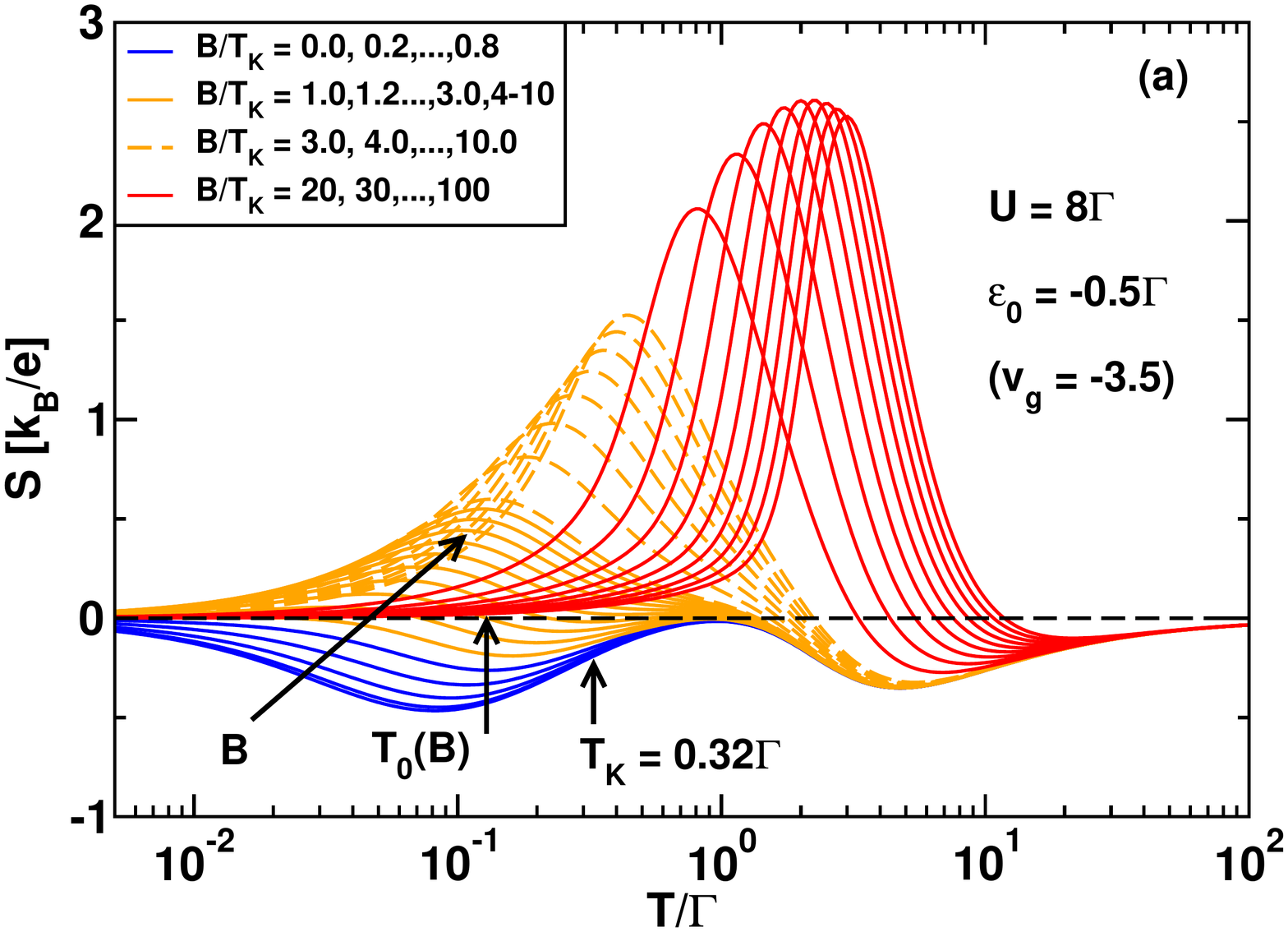}
  \includegraphics[width=0.95\columnwidth]{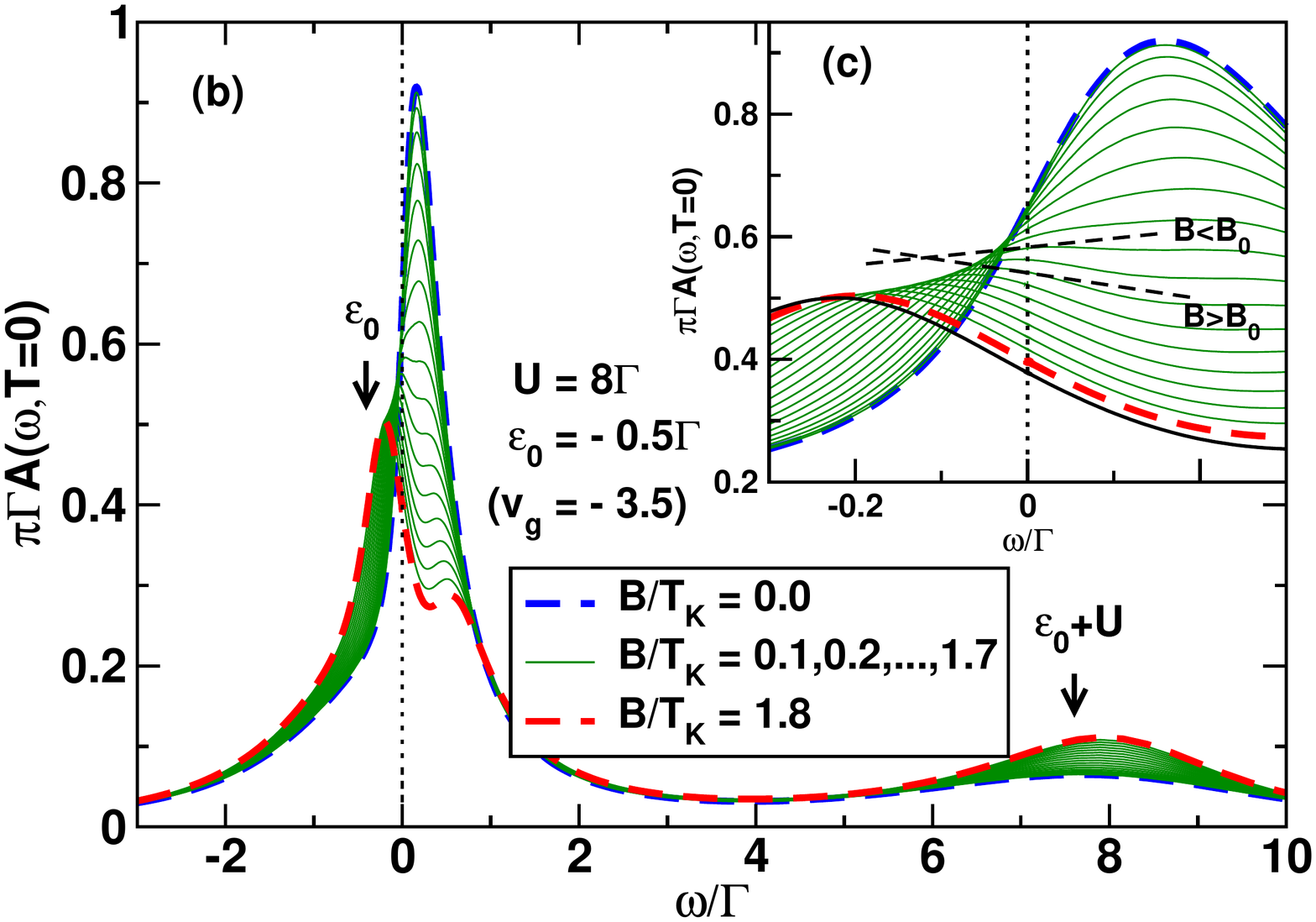}
  \caption 
  {(a) Thermopower $S$  vs 
    temperature $T/\Gamma$, and, (b), spectral function $\pi\Gamma A(\omega,T=0)$ vs 
    $\omega/\Gamma$ for 
    increasing values of the magnetic field $B/T_{\rm K}$ in the mixed valence  regime 
    with $\varepsilon_0=-0.5\Gamma$ and $U=8\Gamma$. 
    Increasing $B$-values in (a) indicated by black arrow. $T_{\rm K}\approx 0.32\Gamma$
    [from Eq.~\ref{eq:tk-spin}], is close to the mixed valence low-energy scale 
    $\Delta=\Gamma/2$. 
    For $B<B_0\approx 0.9T_{\rm K}$, the thermopower 
    exhibits no sign change as a function of $T$ (blue solid lines), while for $B\gtrsim B_0$
    a single sign change at $T_{0}(B)$ occurs. 
    Black vertical arrows in (b): bare Hubbard 
    satellite peaks at $\omega=\varepsilon_{0}$ and $\omega=\varepsilon_0+U$ (for $B=0$). 
    Inset (c): 
    details of the low-energy mixed valence resonance for different magnetic fields. 
    The resonance splits at a field $B_{c}\approx 0.8T_{\rm K}<B_{0}\approx 0.9T_{\rm K}$. 
    Black dashed lines: schematic slope of the spectral function at the Fermi level for $B<B_{0}$ and $B>B_0$. 
    NRG parameters as in Fig.~\ref{fig:fig1}. 
  }
  \label{fig:fig10}
\end{figure*}
 We first recall  that in zero magnetic field, the thermopower, $S(T)$, in the mixed valence
and empty orbital regimes, is of one sign for all $T$ (negative for
${\rm v}_g<0$, positive for ${\rm v}_g>0$) (with the present definition of ${\rm v}_g$) \cite{Costi2010}. This contrasts with the
thermopower in the Kondo regime, which exhibits two characteristic sign changes
at $T=T_1(B=0)\gtrsim T_{\rm K}$ and $T=T_2(B=0)\gtrsim \Gamma$ \cite{Costi2010}.
In the main text we showed that, in the 
presence of a magnetic field $B>B_0$, the thermopower in the Kondo regime exhibits, 
in addition to the sign changes at $T_1(B)$ and $T_2(B)$, a sign change at a low temperature
$T=T_0(B)$. The latter is closely related to the splitting of the asymmetric Kondo
resonance for fields $B>B_c$, with $B_c$ of order $T_{\rm K}$. The
same sign change is present also in the mixed valence regime, as
shown in Fig.~\ref{fig:fig10}(a), and contrasts with the absence of any
sign change in this regime for $B=0$. As in the Kondo case, the sign change in the low temperature $S(T)$ upon increasing $B$ through $B_0$ can be understood as a sign change in the slope of the $T=0$
spectral function at the Fermi level upon increasing the magnetic
field. It correlates approximately with a splitting of the renormalized mixed valence
resonance at a comparable magnetic field [see Fig.~\ref{fig:fig10}(b)].
Note that the mixed valence resonance at $B=0$ is renormalized 
by the Coulomb interaction from its bare position 
at $\omega=\varepsilon_0=-0.5\Gamma$ to a position, 
$\omega=\tilde{\varepsilon}_0>0$, close to, 
but above the Fermi level. This renormalized resonance, having, at $B=0$, a positive slope at
the Fermi level, acquires a negative slope at $B>B_0$ after the resonance has already split at a
somewhat smaller field $B_c$.
\begin{figure*}[t]
  \centering 
  \includegraphics[width=0.95\columnwidth]{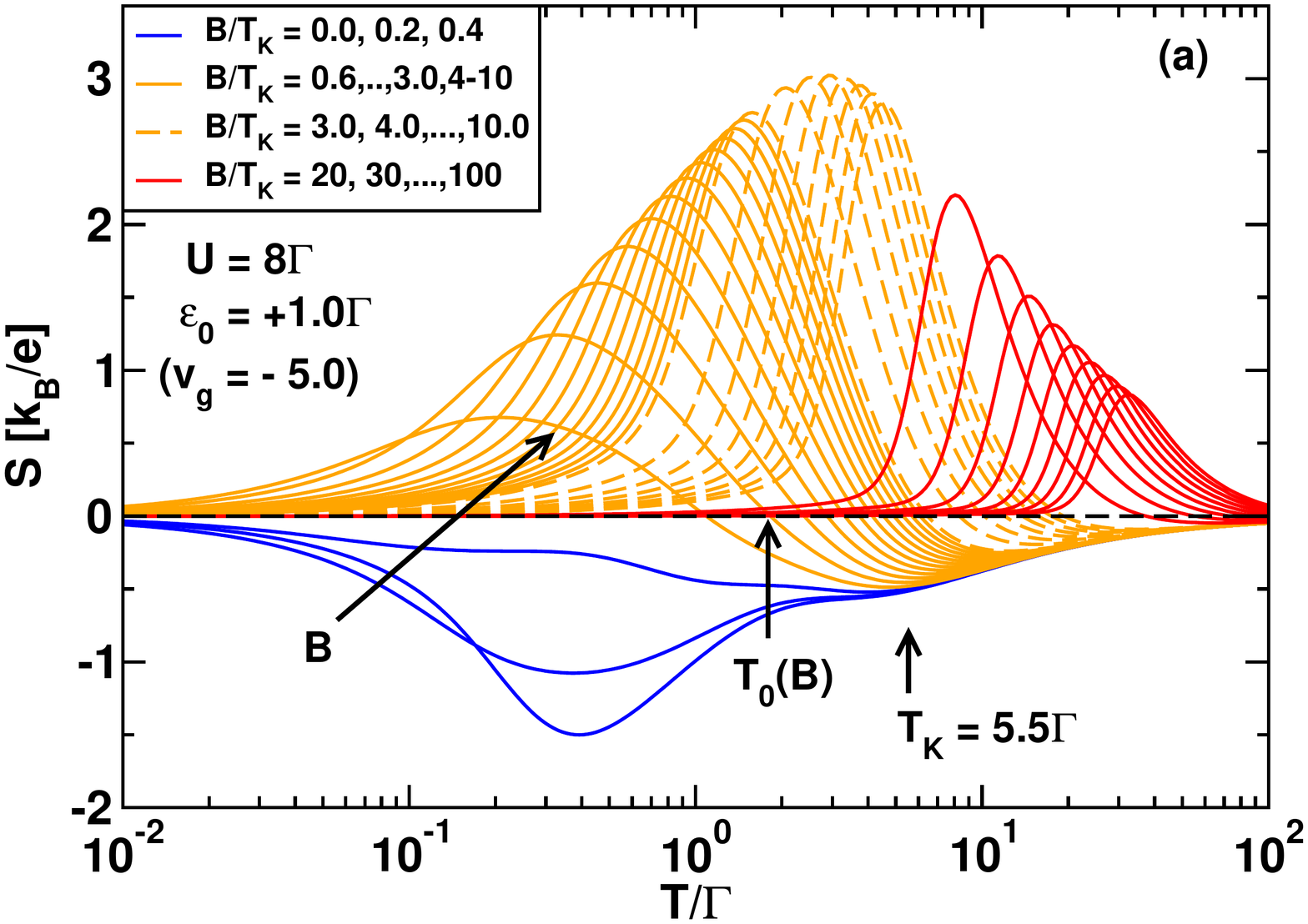}
  \includegraphics[width=0.95\columnwidth]{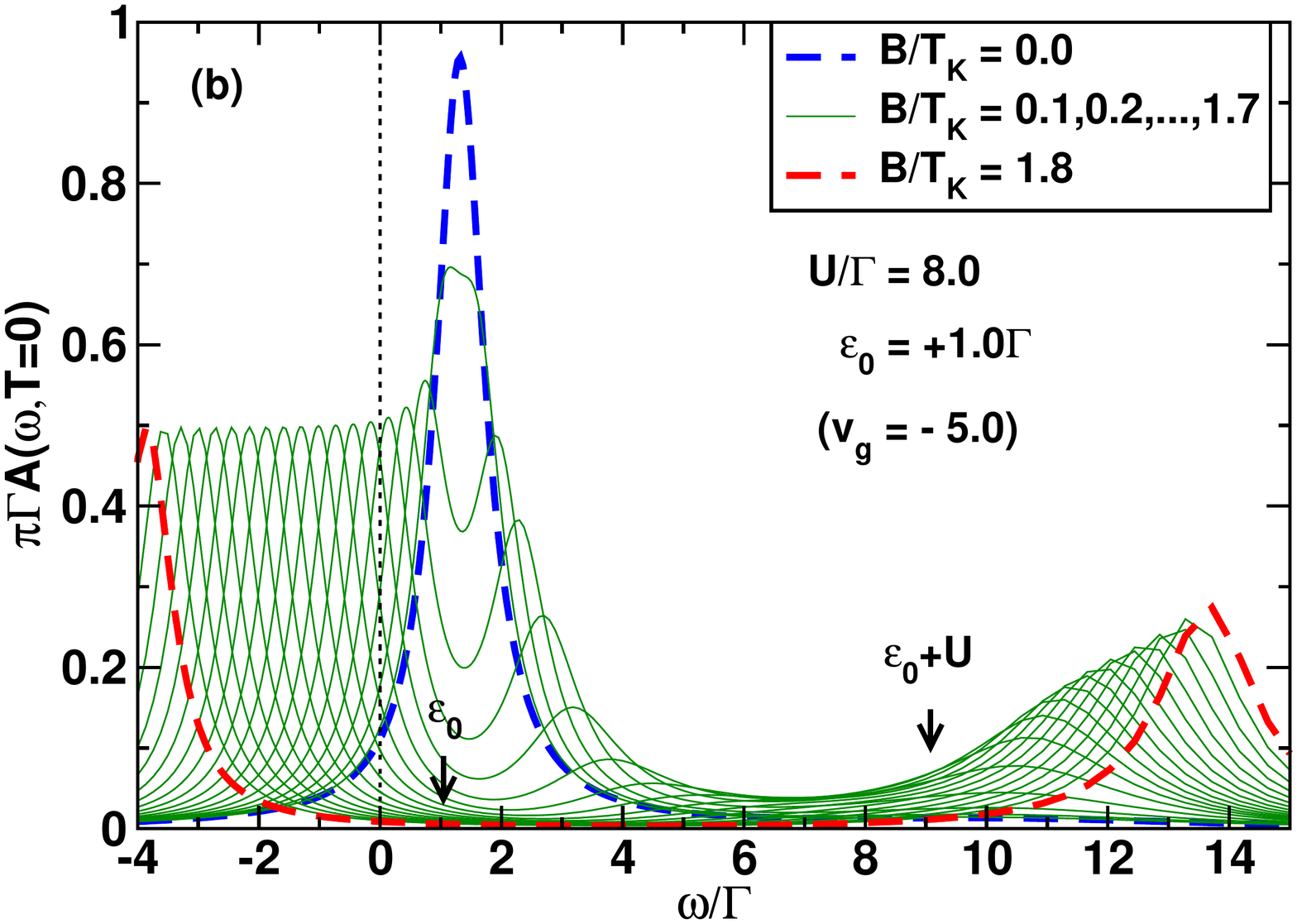}
  \caption 
  {
    (a) Thermopower $S$ vs temperature $T/\Gamma$, and, (b), spectral function $\pi\Gamma A(\omega,T=0)$ vs 
    $\omega/\Gamma$ for increasing values of the magnetic field $B/T_{\rm K}$
    in the empty orbital  regime with $\varepsilon_0=+\Gamma$ and $U=8\Gamma$.
    $T_{\rm K}\approx 5.5\Gamma$ [from Eq.~\ref{eq:tk-spin}], indicated by the vertical 
    arrow in (a), is larger than the true low-energy scale for this
    regime, given by $\varepsilon_0=\Gamma$ (nevertheless, for
    consistency in notation with the Kondo regime, we continue to use $T_{\rm K}$).
    For $B<B_0\approx 0.5T_{\rm K}\approx 2.75\Gamma$ (blue solid lines), the thermopower 
    exhibits no sign change as a function of $T$, while for $B\gtrsim B_0$
    a single sign change at $T_{0}(B)$ occurs.
    Black vertical arrows in (b): bare Hubbard 
    peaks at $\omega=\varepsilon_{0}$ and $\omega=\varepsilon_0+U$ (for $B=0$). 
    The splitting of the resonance at 
    $\omega=\varepsilon_0$,  for fields $B>B_{c}\approx \Gamma$, 
    eventually leads to a change in slope of the spectral function at 
    the Fermi level and hence a sign change in $S(T)$ at low $T$ for $B\gtrsim B_{0}>B_c$. 
    NRG parameters as in Fig.~\ref{fig:fig1}.
      }
  \label{fig:fig11}
\end{figure*}

Similarly, the thermopower in the empty orbital regime, shown in Fig.~\ref{fig:fig11}(a), 
exhibits, for sufficiently large $B>B_0$, a sign change at $T=T_0(B)$. This contrasts with the 
absence of a sign change at $B=0$ in this regime \cite{Costi2010}. The sign change 
for $B>B_0$ correlates with, approximately, the splitting of the resonance at 
$\omega=\varepsilon_0$ in the spectral function for 
sufficiently large $B>B_0>B_c$ [see Fig.~\ref{fig:fig11}(b)]. Thus we see that in both the mixed valence and empty orbital regimes, the sign change of the low-temperature thermopower for $B>B_0$ results from a clear separation of the up- and down-spin components of the spectral function with increasing magnetic field, which eventually changes the slope of the spectral function at the Fermi level for $B>B_0$. In contrast, in the Kondo regime, due to the strong pinning of the Kondo resonance to the immediate vicinity of the Kondo resonance [see Fig.~\ref{fig:fig1}(b)], the above simple picture does not apply. Instead, the effect of a magnetic field is to subtly shift spectral weight from above to below the Fermi level with increasing $B$, such that eventually a sign change in the slope occurs for $B>B_0>B_c$, thereby resulting in a change in sign of the low-temperature thermopower.
\section{$T_1$ and $T_2$ for different $U/\Gamma$}
\label{sec:table}
While values of $U/\Gamma$ of order $3$ are common for semiconductor 
quantum dots exhibiting Kondo physics 
\cite{Goldhaber1998b,Cronenwett1998,Kretinin2011}, $U/\Gamma$ can be 
significantly larger for molecular quantum dots
\cite{Park2002,Roch2008,Dutta2019}. It is therefore of some interest
to give theoretical estimates for the limiting values of $T_1/\Gamma$ and
$T_2/\Gamma$ at midvalley (${\rm v}_g\to 0$) for different $U/\Gamma$
(and for $B=0$).
Table~\ref{Table2} provides this information and lists also
$T_1/T_{\rm K1}$, where $T_{\rm K1}$ is
the commonly used midvalley perturbative Kondo scale. In addition, we 
list the Kondo scale $T_{\rm K}$ defined via the static $T=0$ spin 
susceptibility (evaluated at ${\rm v}_g=0$) given in Eq.~(\ref{eq:tk-spin}),
which is the usual scale used in theoretical works  on
the Kondo problem \cite{Hewson1997}. Note also, that since $T_{1}$ and $T_2$ are weak
functions of gate voltage in the 
Kondo regime (in contrast to the Kondo scale) \cite{Costi2010,Dutta2019}, the midvalley values listed
in Table~\ref{Table2} can be used as rough estimates for $T_1$ and $T_2$ for any gate
voltage in this regime.
For quantum dots with $U/\Gamma\gg 1$, it
would appear from Table~\ref{Table2} that
$T_1$ is inaccessible since $T_1/T_{\rm K1}\gg 1$ at
midvalley. However, for gate voltages ${\rm v}_g$ approaching the
mixed valence regime, this ratio will become smaller, allowing $T_1$
to be accessed experimentally even for quantum dots with $U/\Gamma\gg
1$, as is typically the case for molecular junctions \cite{Roch2008,Dutta2019}. 
\begin{table}[t]
\begin{ruledtabular}
\begin{tabular}{cccccc}
\multicolumn{1}{c}{$U/\Gamma$} & 
\multicolumn{1}{c}{$T_{1}/\Gamma$}&
\multicolumn{1}{c}{$T_{1}/T_{\rm K1}$}& 
\multicolumn{1}{c}{$T_{2}/\Gamma$}& 
\multicolumn{1}{c}{$T_{\rm K1}/\Gamma$}& 
\multicolumn{1}{c}{$T_{\rm K}/\Gamma$}\\
\colrule 
$3$   & $0.1374$ & $1.67$            & $0.6813$  & $8.208\times 10^{-2}$ & $10.56\times 10^{-2}$\\ 
$3.2$& $0.1266$ &$1.75$             & $0.7683$  & $7.245\times 10^{-2}$& $9.192\times 10^{-2}$\\ 
$4$   & $0.0983$ &$2.28$             & $1.1009$  & $4.321\times 10^{-2}$& $5.244\times 10^{-2}$\\ 
$5$   & $0.0762$ &$3.46$             & $1.5041$  & $2.203\times 10^{-2}$& $2.575\times 10^{-2}$\\ 
$6$   & $0.0612$ &$5.56$             & $1.9021$  & $1.100\times 10^{-2}$& $1.254\times 10^{-2}$\\ 
$8$   & $0.0426$ &$16.13$           & $2.6924$  & $2.641\times 10^{-3}$& $2.913\times 10^{-3}$\\ 
$10$ & $0.0322$ &$52.46$           & $3.4800$  & $6.138\times 10^{-4}$& $6.639\times 10^{-4}$\\ 
$12$ & $0.0258$ &$184.55$         & $4.2666$  & $1.398\times 10^{-4}$& $1.492\times 10^{-4}$\\ 
$14$ & $0.0215$ &$685.15$         & $5.0480$  & $3.138\times 10^{-5}$& $3.319\times 10^{-5}$\\ 
$16$ & $0.0185$ &$2652.71$       & $5.8366$  & $6.974\times 10^{-6}$& $7.325\times 10^{-6}$\\ 
$18$ & $0.0162$ &$10533.16 $    & $ 6.6250$ & $1.538\times 10^{-6}$& $1.606\times 10^{-6}$\\ 
$20$ & $0.0145$ &$43029.33$     & $7.4130$  & $3.369\times 10^{-7}$& $3.505\times 10^{-7}$\\ 
$22$ & $0.0131$ &$178304.07$   & $8.1969$  & $7.347\times 10^{-8}$& $7.614\times 10^{-8}$\\ 
$24$ & $0.0119$ &$745983.91$   & $8.9800$  & $1.595\times 10^{-8}$& $1.648\times 10^{-8}$\\ 
$26$ & $0.0109$ &$3158029.08$ & $9.7601$  & $3.451\times 10^{-9}$& $3.557\times 10^{-9}$\\ 
\hline 
\end{tabular}
\end{ruledtabular}
\caption 
{Estimates for $T_{1}/\Gamma$, $T_1/T_{\rm K1}$, $T_2/\Gamma$ at ${\rm v}_g\to 0$
  for quantum dots with different values of $U/\Gamma$. Also indicated 
  are the values of the 
  Kondo scales $T_{\rm K1}$
  and $T_{\rm K}$ at midvalley.  The former is defined by Eq.~(\ref{eq:tk-vs-vg}),  
  evaluated at midvalley (${\rm v}_g=0$), and corresponds to the Kondo  
  scale used in the experiment \cite{Svilans2018}. The latter 
  is defined via the $T=0$ spin susceptibility using 
  Eq.~(\ref{eq:tk-spin}), also evaluated at midvalley. 
  The scale $T_{\rm K}$ is seen to lie within a few percent of $T_{\rm K1}$ for $U/\Gamma \gg 1$. 
}
\label{Table2}
\end{table} 
\section{$B_1/\Gamma$  vs ${\rm v}_g$ for different $U/\Gamma$}
\label{sec:B1Gamma}
\begin{figure}[t]
  \centering 
  \includegraphics[width=0.95\columnwidth]{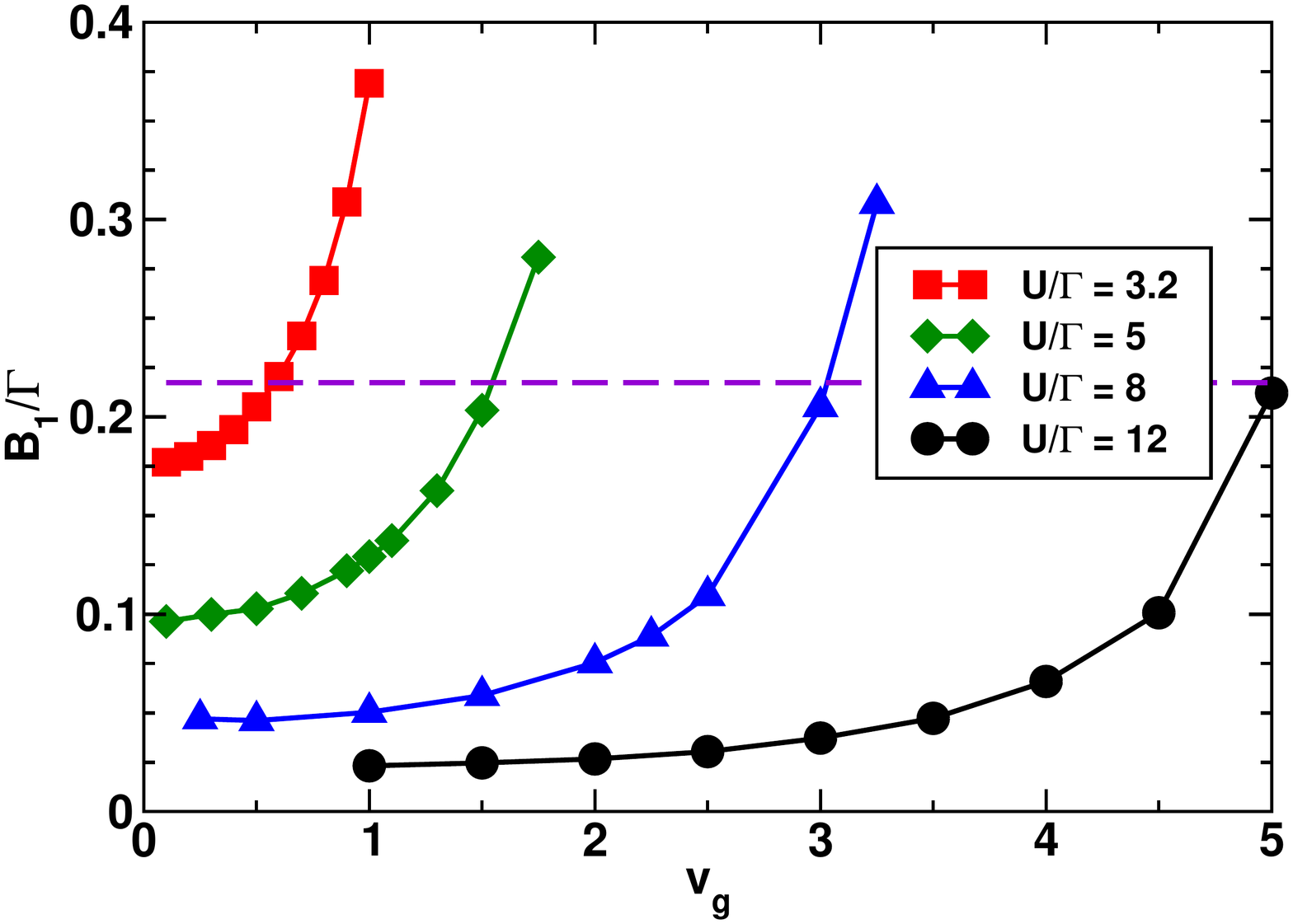}
  \caption  
  {$B_1$ (in units of $\Gamma$) vs ${\rm v}_g$ for different $U/\Gamma$. The horizontal dashed line
    corresponds to the field value $B=0.5 \,{\rm T}$ of relevance in Sec.~\ref{sec:comparison}.
    NRG parameters as in Fig.~\ref{fig:fig1}.
  }
  \label{fig:fig12}
\end{figure}  
 As stated in Sec.~\ref{sec:sign-changes}, the value of $B_1$ in the Kondo regime 
correlates with the scale $T_1$. This can be deduced from  Fig.~\ref{fig:fig12}
which shows $B_1$ in units of $\Gamma$.
The values of $B_1/\Gamma$ at ${\rm v}_g=0$ are seen to correlate
with $T_{1}/\Gamma$ at ${\rm v}_g=0$ from Table~\ref{Table2}:
$B_1/\Gamma\approx 0.14,0.095,0.046$ and $0.024$ as compared to
$T_{1}/\Gamma\approx 0.13,0.08,0.043$ and $0.026$, for
$U/\Gamma=3.2,5,8$ and $12$, respectively. 
\section{Comparison between $S$ and linear-response thermocurrent $I_{\rm th}/\Delta T =GS$}
\label{sec:GS-comparison}
\begin{figure*}[t]
\centering 
\includegraphics[width=0.95\columnwidth]{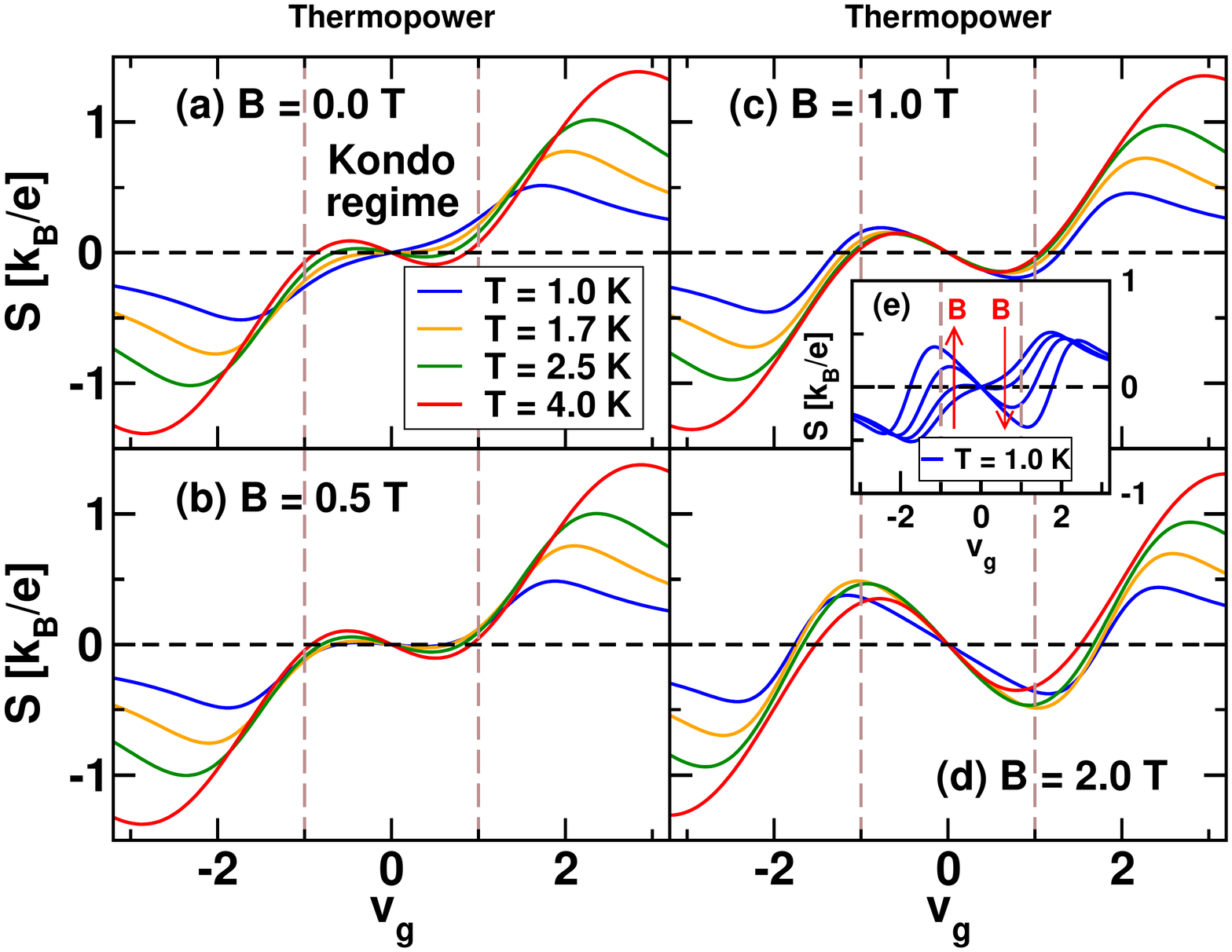}\hspace{1cm}
  \includegraphics[width=0.95\columnwidth]{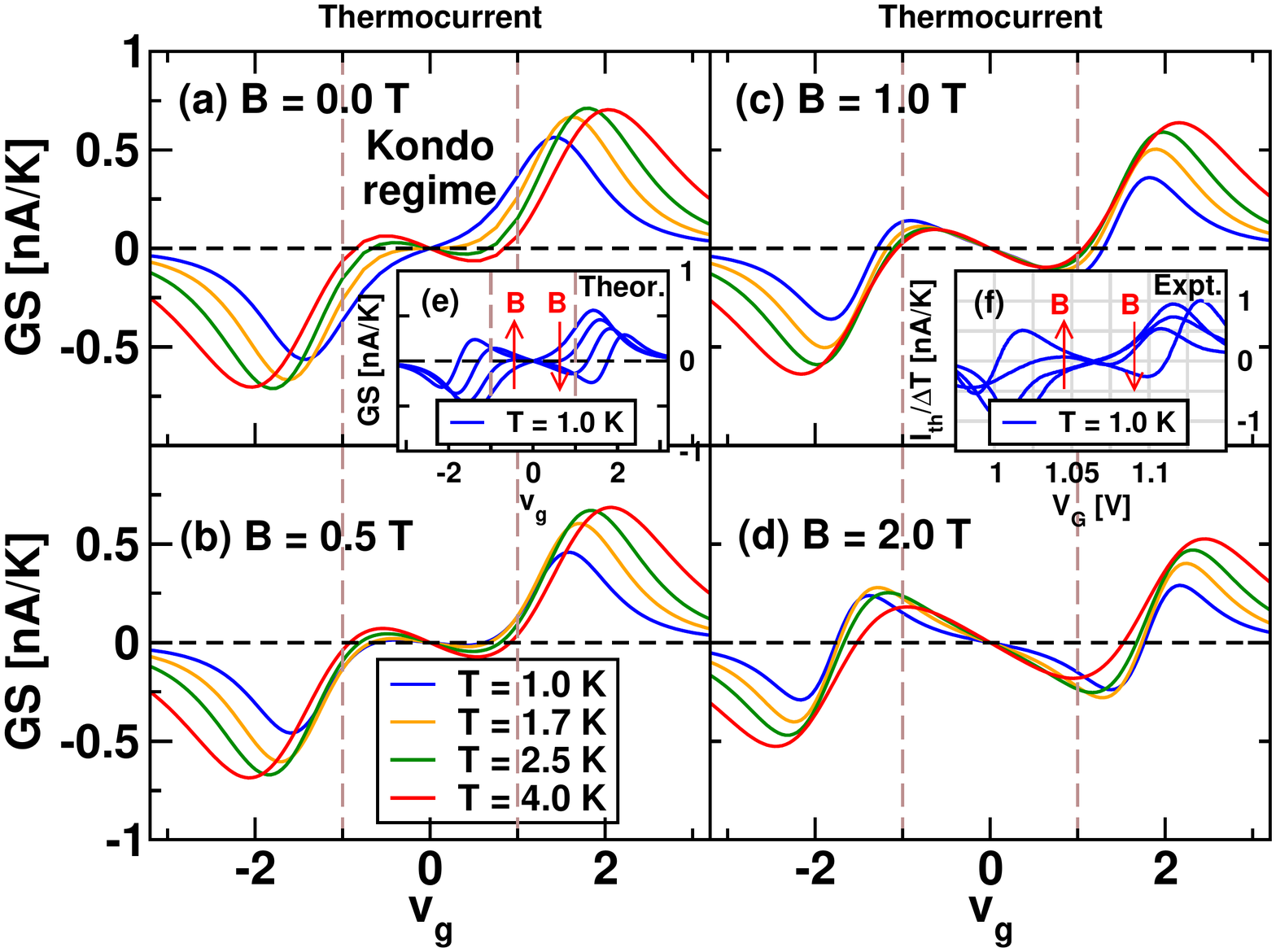}
  \caption{
Left panels (a)-(d): calculated $S$ vs ${\rm v}_g$ at the temperatures $T$ and the four magnetic fields $B$ of the experiment \cite{Svilans2018}. 
Vertical dashed lines delineate Kondo ($|{\rm v}_g|\lesssim 1.0$) from mixed valence and empty (full) orbital regimes at larger $|{\rm v}_g|$. 
  Inset (e): lowest temperature data ($T=1.0\, {\rm K}$) from (a)-(d) for increasing $B$ (red arrows). 
  Right panels (a)-(d):  the calculated linear-response thermocurrent $I_{\rm th}/\Delta T = GS$ vs gate-voltage ${\rm v}_g$
  at the four magnetic fields and temperatures of the experiment \cite{Svilans2018}). 
  Inset (e): calculated thermocurrent ($GS$) at the lowest temperature ($T=1.0\, {\rm K}$) from (a)-(d) for increasing $B$ (red arrows).
  Inset (f): experimental   thermocurrent ($I_{th}/\Delta T$) at the lowest temperature ($T=1.0\, {\rm K}$) with data taken from Fig.~\ref{fig:fig6} (right panels) of the experiment \cite{Svilans2018}.
  NRG parameters as in Fig.~\ref{fig:fig1}.}
\label{fig:fig13}
\end{figure*}
While the linear-response thermopower $S(T)$ and thermocurrent  $I_{\rm th}/\Delta T=GS$ [see discussion following Eq.~(\ref{eq:linear-response}) in Sec.~\ref{sec:comparison}] exhibit the same sign changes at $T_0(B),T_1(B)$ and $T_2(B)$ as discussed in the main text, their gate-voltage dependence at different fields exhibits some qualitative differences which we would like to mention in the context of the
  experiment \cite{Svilans2018}. Figure~\ref{fig:fig13} compares the calculated gate-voltage dependence of the thermopower $S$ (left panels) with that of the linear-response thermocurrent $I_{\rm th}/\Delta T=GS$ (right panels) for the temperature and field values of the experiment \cite{Svilans2018}. 
We note the stronger reduction of the thermocurrent $GS$ at large fields ($B=2.0\,T$) in the Kondo regime [Fig.~\ref{fig:fig13}(d) (right panels)] as compared to that in the thermopower $S$ [Fig.~\ref{fig:fig13}(d) (left panels)]. This
reflects the strong suppression with field of the Kondo resonance, and hence of $G$ and $GS$, particularly at low temperatures (blue curves). In contrast, the thermopower $S$, for $|{\rm v}_g|>0$, after an initial suppression with increasing field from $B=0.0\,T$ to $B=0.5\,T$ [Figs.~\ref{fig:fig13}(a) and \ref{fig:fig13}(b) (left panels)],
starts to increase with increasing field [$B=1.0\,$ and $2.0\, T$ curves in [Figs.~\ref{fig:fig13}(c) and \ref{fig:fig13}(d) (left panels)].
This latter effect is due to the fact that a magnetic field makes the total spectral function more asymmetric (when $|{\rm v}_g|>0$) \cite{Costi2001,Hofstetter2001} and thereby leads to an enhancement of $S$ for sufficiently large $B$ (since $S$ measures the asymmetry of the spectral function about the Fermi level).

The insets Fig.~\ref{fig:fig13}(e) for $S$ (left panels)  and $GS$ (right panels) demonstrate, as mentioned in Sec.~\ref{sec:sign-changes}, another feature of the thermopower of Kondo-correlated quantum dots, namely, that for $T\lesssim T_1$, $S(T)$ [$G(T)S(T)$] is of opposite sign for $B\to 0$ (here $B=0 T$) and $B>B_1$ (here, $B=1.0 T$ and $2.0 T$), which is also consistent with the experiment \cite{Svilans2018}, as can be seen in the comparison between theory [Fig.~\ref{fig:fig13}(e), right panels] and experiment [Fig.~\ref{fig:fig13}(f), right panels] for the field dependence of the
thermocurrent at the lowest temperature $T=1.0\,K$ [using the experimental data 
from Fig.~\ref{fig:fig6} (right panels)].
\bibliography{mtep} 
\end{document}